\documentclass{emulateapj}

\usepackage{lineno}
\usepackage{graphicx}
\usepackage{auto-pst-pdf}
\usepackage{multirow}
\usepackage{natbib}
\usepackage{color}
\usepackage[normalem]{ulem}
\makeatletter

\newcommand{\Rmnum}[1]{\expandafter\@slowromancap\romannumeral #1@}
\makeatother

\begin{document}

\title{Deep morphological and spectral study of the SNR RCW 86 with \textit{Fermi}-LAT}
\author{
M.~Ajello\altaffilmark{2}, 
L.~Baldini\altaffilmark{3,4}, 
G.~Barbiellini\altaffilmark{5,6}, 
D.~Bastieri\altaffilmark{7,8}, 
R.~Bellazzini\altaffilmark{9}, 
E.~Bissaldi\altaffilmark{10}, 
E.~D.~Bloom\altaffilmark{4}, 
R.~Bonino\altaffilmark{11,12}, 
E.~Bottacini\altaffilmark{4}, 
T.~J.~Brandt\altaffilmark{13}
% ...
}
% ...
\altaffiltext{2}{Department of Physics and Astronomy, Clemson University, Kinard Lab of Physics, Clemson, SC 29634-0978, USA}
\altaffiltext{3}{Universit\`a di Pisa and Istituto Nazionale di Fisica Nucleare, Sezione di Pisa I-56127 Pisa, Italy}
\altaffiltext{4}{W. W. Hansen Experimental Physics Laboratory, Kavli Institute for Particle Astrophysics and Cosmology, Department of Physics and SLAC National Accelerator Laboratory, Stanford University, Stanford, CA 94305, USA}
\altaffiltext{5}{Istituto Nazionale di Fisica Nucleare, Sezione di Trieste, I-34127 Trieste, Italy}
\altaffiltext{6}{Dipartimento di Fisica, Universit\`a di Trieste, I-34127 Trieste, Italy}
\altaffiltext{7}{Istituto Nazionale di Fisica Nucleare, Sezione di Padova, I-35131 Padova, Italy}
\altaffiltext{8}{Dipartimento di Fisica e Astronomia ``G. Galilei'', Universit\`a di Padova, I-35131 Padova, Italy}
\altaffiltext{9}{Istituto Nazionale di Fisica Nucleare, Sezione di Pisa, I-56127 Pisa, Italy}
\altaffiltext{10}{Istituto Nazionale di Fisica Nucleare, Sezione di Bari, I-70126 Bari, Italy}
\altaffiltext{11}{Istituto Nazionale di Fisica Nucleare, Sezione di Torino, I-10125 Torino, Italy}
\altaffiltext{12}{Dipartimento di Fisica Generale ``Amadeo Avogadro'' , Universit\`a degli Studi di Torino, I-10125 Torino, Italy}
\altaffiltext{13}{NASA Goddard Space Flight Center, Greenbelt, MD 20771, USA}
% ...


\begin{abstract}

RCW 86 is a young supernova remnant (SNR) showing a shell-type structure at several wavelengths and is thought to be an efficient cosmic-ray (CR) accelerator. Earlier \textit{Fermi} Large Area Telescope results reported the detection of $\gamma$-ray emission coincident with the position of RCW 86 but its origin (leptonic or hadronic) remained unclear due to the poor statistics. Thanks to 6.5 years of data acquired by the \textit{Fermi}-LAT and the new event reconstruction Pass 8, we report the significant detection of spatially extended emission coming from RCW 86. The spectrum is described by a power-law function with a very hard photon index ($\Gamma = 1.42 \pm 0.1_{\rm stat} \pm 0.06_{\rm syst}$) in the 0.1--500 GeV range and an energy flux above 100 MeV of ($2.91$ $\pm$ $0.8_{\rm stat}$ $\pm$ $0.12_{\rm syst}$) $\times$ $10^{-11}$ erg cm$^{-2}$ s$^{-1}$. Gathering all the available multiwavelength (MWL) data, we perform a broadband modeling of the nonthermal emission of RCW 86 to constrain parameters of the nearby medium and bring new hints about the origin of the $\gamma$-ray emission. For the whole SNR, the modeling favors a leptonic scenario in the framework of a two-zone model with an average magnetic field of 10.2 $\pm$ 0.7 $\mu$G and a limit on the maximum energy injected into protons of 2 $\times$ 10$^{49}$ erg for a density of 1 cm$^{-3}$. In addition, parameter values are derived for the North-East (NE) and South-West (SW) regions of RCW 86, providing the first indication of a higher magnetic field in the SW region.

\end{abstract}

\maketitle

%%%========================================================================================================%%%
\section{Introduction}\label{sec:introduction}

\setcounter{footnote}{0}

It is widely believed that supernova remnants (SNRs) are the primary sources of Galactic cosmic rays (CRs) observed on Earth, up to the knee energy at $\sim10^{15}$~eV,  as first proposed by \cite{Ginzburg1961}. CRs can be accelerated to very high energies at collisionless shocks driven by Supernova (SN) explosions through diffusive shock acceleration \citep[DSA;][]{Axford1977, Krymskii1977, Bell1978a,Bell1978b, Blandford1978}. During this process the kinetic energy released in SN explosions has to be transferred to CRs with an efficiency of $\sim$ 10\% \citep{Ginzburg1964}. A description of the acceleration process can be achieved within a non-linear DSA theory \citep{Malkov2001} with magnetic field amplification, probably by accelerated particles themselves, due to streaming instability \citep{Bell2004, Amato2006, Caprioli2008}. The strong evidence for large magnetic fields in the shock region is given by the observation of narrow filaments of non-thermal X-ray radiation in young SNRs \citep{Ballet2006, Vink2012}. The non-thermal emission produced through the interaction of accelerated particles with radiation and/or matter in the environment of the SNR, via synchrotron (SC), inverse Compton (IC), non-thermal bremsstrahlung, hadronic interactions and subsequent $\pi^0$ decay, gives information about the particle acceleration mechanisms at work in these sources.  

RCW 86 \citep{Rodgers1960}, also known as MSH 14$-$6{\em3} \citep{Mills1961} or G315.4-2.3, is a SNR located in the southern sky. The origin of this SNR is still debated but recent studies \citep{Williams2011, Broersen2014} suggest that RCW 86 is associated to the historical SN 185 \citep{Stephenson2002} and is the result of a Type Ia explosion, also supported by the large amount of Fe \cite[$\sim$1 M$_{\odot}$;][]{Yamaguchi2011}. A large shell ($\sim 40'$ in diameter) is clearly detected in radio \citep{Kesteven1987}, optical \citep{Smith1997}, infrared \citep{Williams2011}, X-rays \citep{Pisarski1984} and very-high-energy (VHE; E $>$ 0.1 TeV) $\gamma$-rays \citep{Aharonian2009, Abramowski2015}. At high-energy (HE, 0.1 $<$ E $<$ 100 GeV) $\gamma$-rays, \cite{Lemoine2012} derived upper limits on the flux and \cite{Yuan2014} reported the detection of a pointlike $\gamma$-ray source matching the position of RCW 86. In 2015, RCW 86 was reported for the first time as an extended source (with a radius of $0\fdg27$ above 50 GeV) in the Second Catalog of Hard \textit{Fermi}-LAT Sources \citep{2FHL2015}. 

X-ray observations of RCW 86 reveal a non-spherically symmetric shell with both thermal ($0.5-2$ keV) and non-thermal ($2-5$ keV) emission, with different morphologies. The soft X-rays are related to optical emission from non-radiative shocks and IR emission from collisionally heated dust, whereas the hard X-ray continuum, located mostly in the southwestern part of the remnant, is due to SC radiation coming from electrons accelerated at the reverse shock of the remnant, as suggested by its spatial correlation with the strong Fe$-$K line emission \citep{Rho2002}. Using \emph{Suzaku} telescope data, \cite{Ueno2007} produced a map of the Fe$-$K line emission in the southwestern part of the remnant, showing that the Fe$-$K line emission correlates well with the radio SC emission. Furthermore, the higher temperature plasma, which mostly contains the strong Fe$-$K line emission, suggests that this line originates from Fe-rich ejecta heated by a reverse shock \citep{Ueno2007, Yamaguchi2008}. This X-ray SC radiation is produced by TeV electrons accelerated at the shock as confirmed by the VHE $\gamma$-ray emission detected with the H.E.S.S. experiment \citep{Aharonian2009, Abramowski2015}. 

The distance of RCW 86 is estimated to be 2.5 $\pm$ 0.5 kpc through the recent proper motion measurements by \cite{Helder2013}, combined with plasma temperature measurements based on the broad H$\alpha$ lines. The ambient density around RCW 86 is inhomogeneous and the shock speed value, as well as the magnetic field, change along the shell-like structure. In particular, in the southwest and northwest regions shocks are slow, around $\sim$ 600--800 km s$^{-1}$ \citep{Long1990, Ghavamian2001}, and post-shock densities are relatively high \cite[$\sim$ 2 cm$^{-3}$;][]{Williams2011}. Whereas, faster shocks \cite[$\sim$ 2700 kms$^{-1}$ and 6000 $\pm$ 2800 km s$^{-1}$;][]{Vink2006,Helder2009} and lower densities \cite[$\sim$ 0.1 -- 0.3 cm$^{-3}$;][]{Yamaguchi2008} have been measured in the northeast (NE) region. The large size of this young remnant as well as the asymmetry in its morphology can be explained by an off-center explosion in a low-density cavity, as proposed by \cite{Williams2011}.

Here we present the results of a deep morphological analysis with the new \textit{Fermi}-LAT event reconstruction set, Pass 8, as well as a study of the broadband emission using the available information in the radio, X-ray and VHE $\gamma$-ray domains.

%%%========================================================================================================%%%
\section{\textit{Fermi}-LAT and Pass 8 description}

The \textit{Fermi}-LAT is a $\gamma$-ray telescope which detects photons by conversion into electron--positron pairs in the energy range between 20 MeV to higher than 500 GeV, as described in \cite{Atwood2009}. The LAT is made of a high-resolution converter/tracker (for direction measurement of the incident $\gamma$-rays), a CsI(Tl) crystal calorimeter (for energy measurement), and an anti-coincidence detector to identify the background by charged particles. The LAT has a large effective area ($\sim$ 8200 cm$^{2}$ on-axis above 1~GeV), a wide field of view ($\sim$ 2.4 sr) as well as good angular resolution (with a 68$\%$ containment radius of $\sim 0\fdg8$ at 1 GeV).

Since the launch of the spacecraft in June 2008, the LAT event-level analysis has been periodically upgraded to take advantage of the increasing knowledge of the \textit{Fermi}-LAT functioning as well as the environment in which it operates. Following Pass 7, released in August 2011, Pass 8 is the latest version of the \textit{Fermi}-LAT data\footnote{\textit{Passes} correspond to the release of upgraded versions of the LAT event-level analysis framework.}. The development of Pass 8 was the result of a long-term effort aimed at a radical revision of the entire event-level analysis and tends to realize the full scientific potential of the LAT \citep{Atwood2013}. Combining the improvement of the effective area, the point-spread function and the energy resolution with the large amount of data collected by the LAT since its launch, Pass 8 is a powerful tool to identify and study extended $\gamma$-ray sources.

%%%========================================================================================================%%%
\section{\textit{Fermi}-LAT observations and data analysis}

We analysed 6.5 years of data collected between August 4$^{th}$, 2008 and January 31$^{st}$, 2015 within a $15^\circ \times 15^\circ$ region centered on the position of RCW 86. We used events with energies between 100 MeV and 500 GeV with a maximum zenith angle of $100^\circ$ to limit the contamination due to the Earth Limb. To assure good quality events, we excluded the time intervals when the \textit{Fermi} spacecraft was within crossed the South Atlantic Anomaly were excluded. We used the version \verb+10-00-03+ of the ScienceTools and the \verb+P8R2_V6+ Instrument Response Functions (IRFs) with the event class \verb+SOURCE+, which corresponds to the best compromise between the number of selected photons and the charged particle residual background for the study of point-like or slightly extended sources.

Two tools were used for the analysis: \verb+gtlike+ for the spectral analysis and \verb+pointlike+ for the spatial analysis. \verb+gtlike+ is a binned maximum likelihood method \citep{Mattox1996} implemented in the \textit{Fermi} Science Tools. \verb+pointlike+ is an alternative code used for fast analysis of \textit{Fermi}-LAT data and able to characterize the extension of a source \citep{Kerr2011}. These tools fit a source model to the data along with models for the residual charged particles and diffuse $\gamma$-ray emission. The Galactic diffuse emission was modeled by the standard LAT diffuse emission ring-hybrid model \verb+gll_iem_v06.fits+ and the residual background and extragalactic radiation were described by a single isotropic component with the spectral shape in the tabulated model \verb+iso_P8R2_SOURCE_V6_v06.txt+. The models are available from the \textit{Fermi} Science Support Center. Sources located in the $15^\circ \times 15^\circ$ region centered on RCW 86 and included in the \textit{Fermi}-LAT Third Source Catalogue \cite[][hereafter 3FGL]{Acero2015}, based on the first four years of Pass 7 data, were added to our spectral-spatial model of the region. Only sources within 5$^\circ$ around the position of RCW 86 were refitted, in addition to the Galactic diffuse and isotropic emission. The energy dispersion, which is defined in terms of the fractional difference between the reconstructed energy and the true energy of the events, was taken into account in both spatial and spectral analysis to consider the imperfection of the energy reconstruction.

\subsection{Morphological analysis}

\begin{figure}[ht]
  \centering
  \includegraphics[width=\columnwidth]{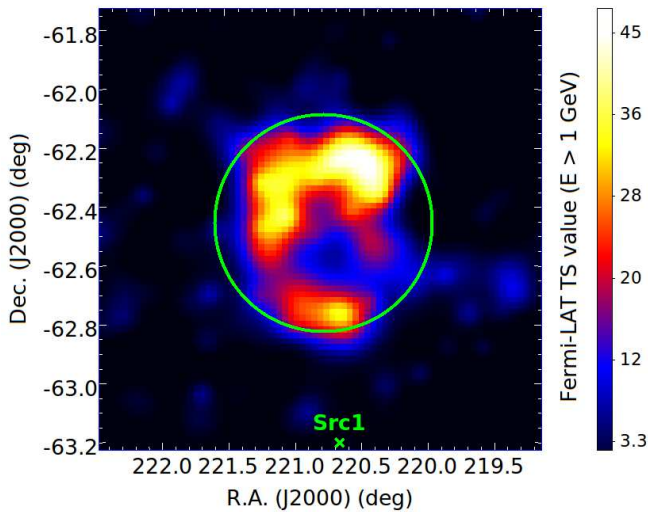}
  \caption{Test Statistic ($TS$) map above 1 GeV centered on RCW 86. The green cross indicates a point-like source that has been added to the source model to fit background emissions in the vinicity of RCW 86. \label{fig:tsmap}}
\end{figure}

As it is critical to have a better angular resolution for the morphological analysis, we perform the spatial analysis above 1 GeV. The resulting PSF is $\sim 0\fdg27$ for a photon index of 1.5 (see Section 3.2). We first fitted the $15^\circ \times 15^\circ$ region centered on RCW 86 with the 3FGL sources, the Galactic diffuse and isotropic emission templates and computed the \textit{Fermi}-LAT Test Statistic ($TS$) map ($4^\circ \times 4^\circ$ with 0.02 degrees per pixel, see Figure~\ref{fig:tsmap}). The $TS$ is defined as twice the difference between the log-likelihood $L_1$ obtained by fitting a source model plus the background model to the data, and the log-likelihood $L_0$ obtained by fitting the background model only, i.e $TS = 2(L_1-L_0)$. Figure~\ref{fig:tsmap} contains the $TS$ value for a point source of fixed photon index $\Gamma = 2$ in each pixel of the map, thus giving a measure of the statistical significance for the detection of a $\gamma$-ray source with that spectrum in excess of the background. The $TS$ map revealed a significant emission coincident with the position of RCW 86, coming essentially from the NE region of the remnant (there is no significant variation of the background emission within $0\fdg5$). Additional excess $\gamma$-ray emissions were also detected in the region of interest. Therefore, we added 4 new sources, denoted with the identifiers Src1, Src2, Src3, and Src4 (only Src1 is visible in the field of view of Figure~\ref{fig:tsmap}), in our model to take into account these signals. These regions of $\gamma$-ray emission are considered to belong to background sources not detected in 3FGL due to the larger dataset and the increased effective area of Pass 8 data.

To determine the best morphology of RCW 86, the data were fitted with different spatial models (point-source, disk, ring) while fitting the spectrum of the source (normalization and spectral index) simultaneously. The results of this first analysis are reported in Table~\ref{tab:results_ptlike}. The significance of the extension of RCW 86 is quantified with $TS_{\rm ext}$, which is defined as twice the difference between the log-likelihood of an extended source model and the log-likelihood of a point-like source model. For the uniform disk hypothesis, the source is significantly extended ($TS_{\rm ext}$ = 68, which corresponds to a significance of $\sim 8\sigma$ for the extension) with respect to the LAT point-spread function (PSF). The fitted radius, $0\fdg37 \pm 0\fdg02_{\rm stat}$, is in good agreement with the size of the SNR as seen in radio \citep{Kesteven1987}, infrared \citep{Williams2011}, X-rays \citep{Pisarski1984} and VHE $\gamma$-rays \citep{Abramowski2015}. Using a ring as a spatial model, the log-likelihood is improved in comparison to that obtained with a uniform disk by only 2.6$\sigma$, which is not enough to claim that the source has a shell-like morphology. The spatial analysis was confirmed by \verb+gtlike+, as shown in Table~\ref{tab:results_gtlike}.

%\begin{deluxetable*}{lcccccc}
%  \tablecaption{Centroid and extension fits of the LAT data using pointlike above 1 GeV. Uncertainties are statistical errors (68\% containement). N$_{\rm dof}$ corresponds to the number of degrees of freedom for each model.\label{tab:results_ptlike}}
%  \tablehead{ 
%    \colhead{Spatial Model} & \colhead{$TS_{\rm ext}$} & \colhead{R.A. ($^\circ$)} & \colhead{Dec. ($^\circ$) } & Radius ($^\circ$) & Inner Radius ($^\circ$) & N$_{\rm dof}$
%  }
%  \startdata
%  Point Source 	& -  & 220.56 $\pm$ 0.07 & $-$62.25 $\pm$ 0.02 & - & - & 3\\
%  Disk		& 68 & 220.73 $\pm$ 0.04 & $-$62.47 $\pm$ 0.03 & $0.37 \pm 0.02$ & -  & 4\\
%  Ring            & 75 & 220.74 $\pm$ 0.02 & $-$62.51 $\pm$ 0.02 & $0.31 \pm 0.02$ & $0.21 \pm 0.02$ & 5\\
%  \enddata
%\end{deluxetable*}

\begin{table*}[ht]
  \caption{Centroid and extension fits of the LAT data using pointlike above 1 GeV. Uncertainties are statistical errors (68\% containment). N$_{\rm dof}$ corresponds to the number of degrees of freedom for each model.\label{tab:results_ptlike}}
  \centering
  \begin{tabular}{lcccccc}
    \hline
    \hline
    %\colhead{Source Name} & \colhead{R.A. ($^\circ$)} & \colhead{Dec. ($^\circ$)} & \colhead{Spectral Index} & \colhead{Flux ($\times$10$^{-12}$ erg/cm$^{2}$/s)} & \colhead{$TS$}
    Spatial Model & $TS_{\rm ext}$ & R.A. ($^\circ$) & Dec. ($^\circ$) & Radius ($^\circ$) & Inner Radius ($^\circ$) & N$_{\rm\ dof}$\rule[-4pt]{0pt}{12pt}\\
    \hline
    Point Source & -  & 220.56 $\pm$ 0.07 & $-$62.25 $\pm$ 0.02 & - & - & 3\rule[-3pt]{0pt}{12pt}\\
    Disk	 & 68 & 220.73 $\pm$ 0.04 & $-$62.47 $\pm$ 0.03 & $0.37 \pm 0.02$ & -  & 4\rule[-3pt]{0pt}{12pt}\\
    Ring         & 75 & 220.74 $\pm$ 0.02 & $-$62.51 $\pm$ 0.02 & $0.31 \pm 0.02$ & $0.21 \pm 0.02$ & 5\rule[-3pt]{0pt}{12pt}\\
    \hline
  \end{tabular}
\end{table*}

\begin{figure}
  \centering
  \begin{tabular}{cc}
      \includegraphics[height=1.25in]{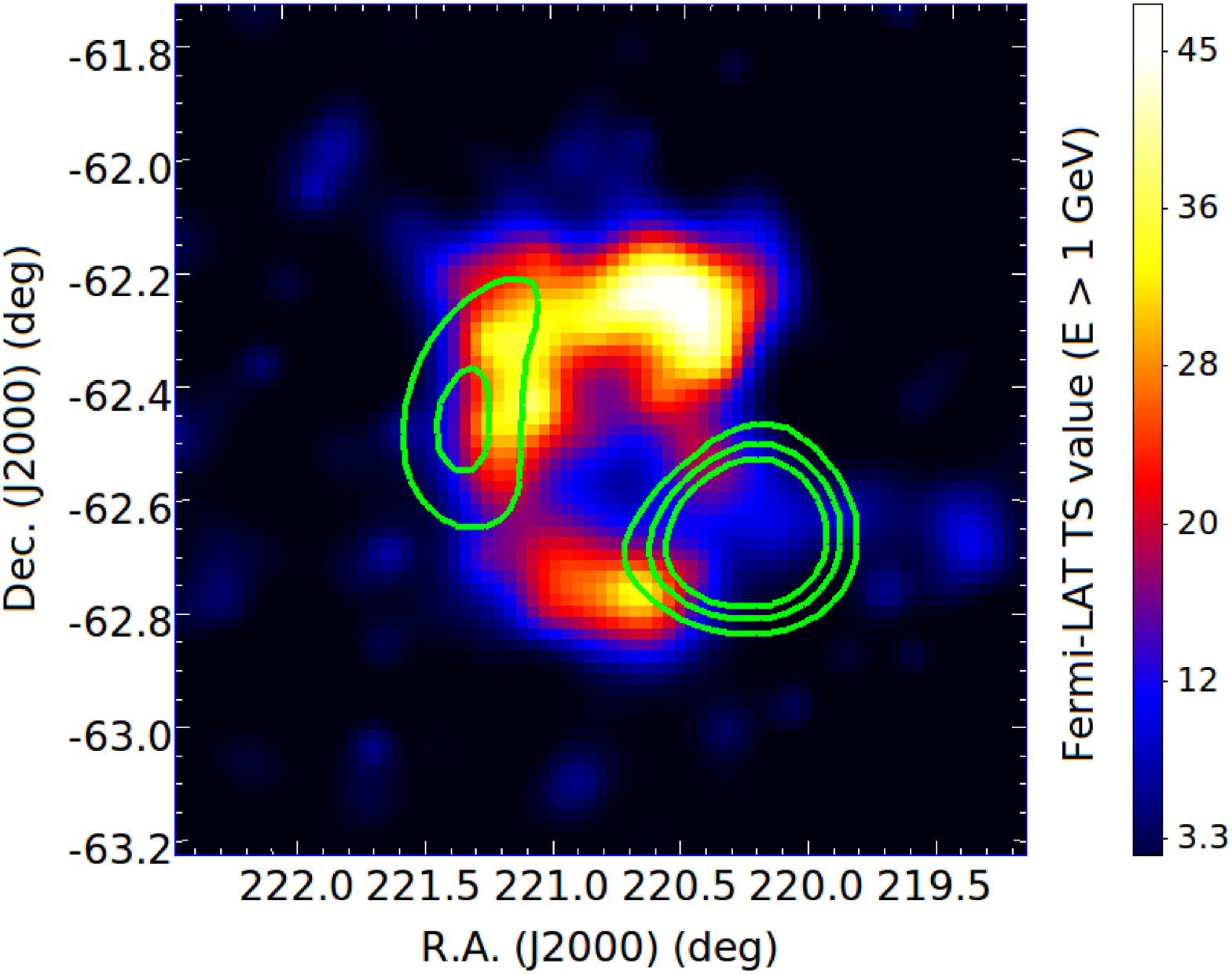}      & \includegraphics[height=1.25in]{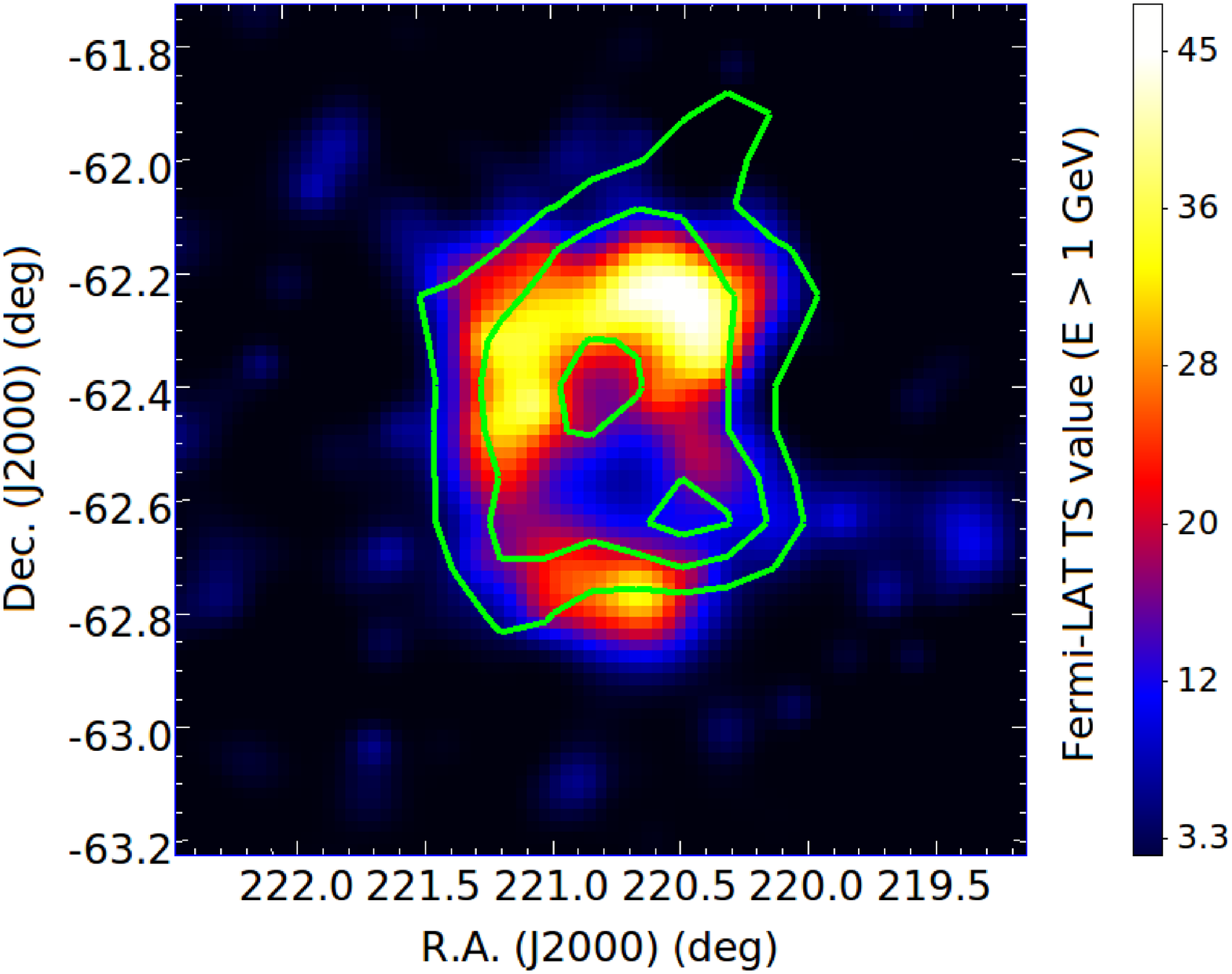}\\
      \includegraphics[height=1.25in]{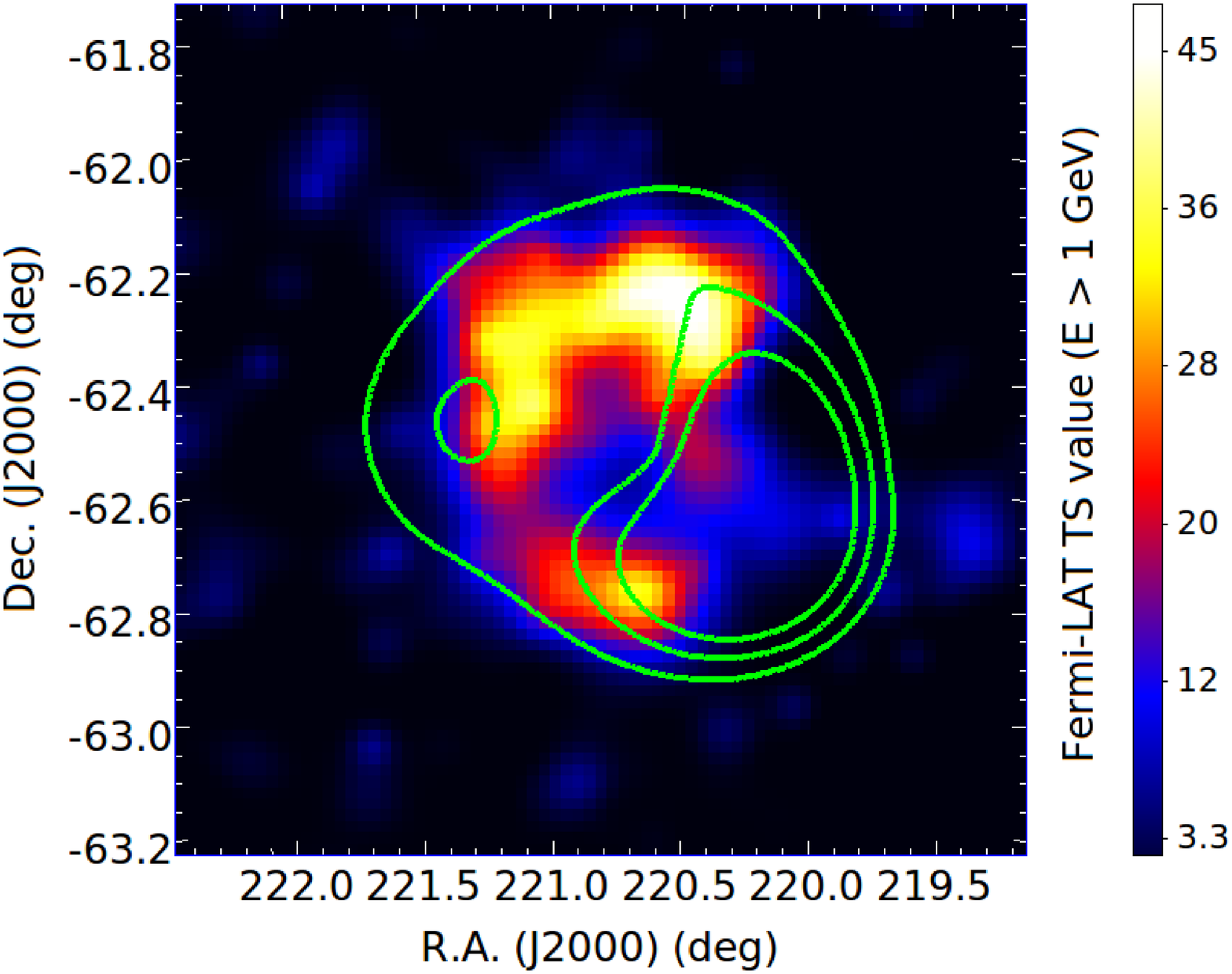} & \includegraphics[height=1.25in]{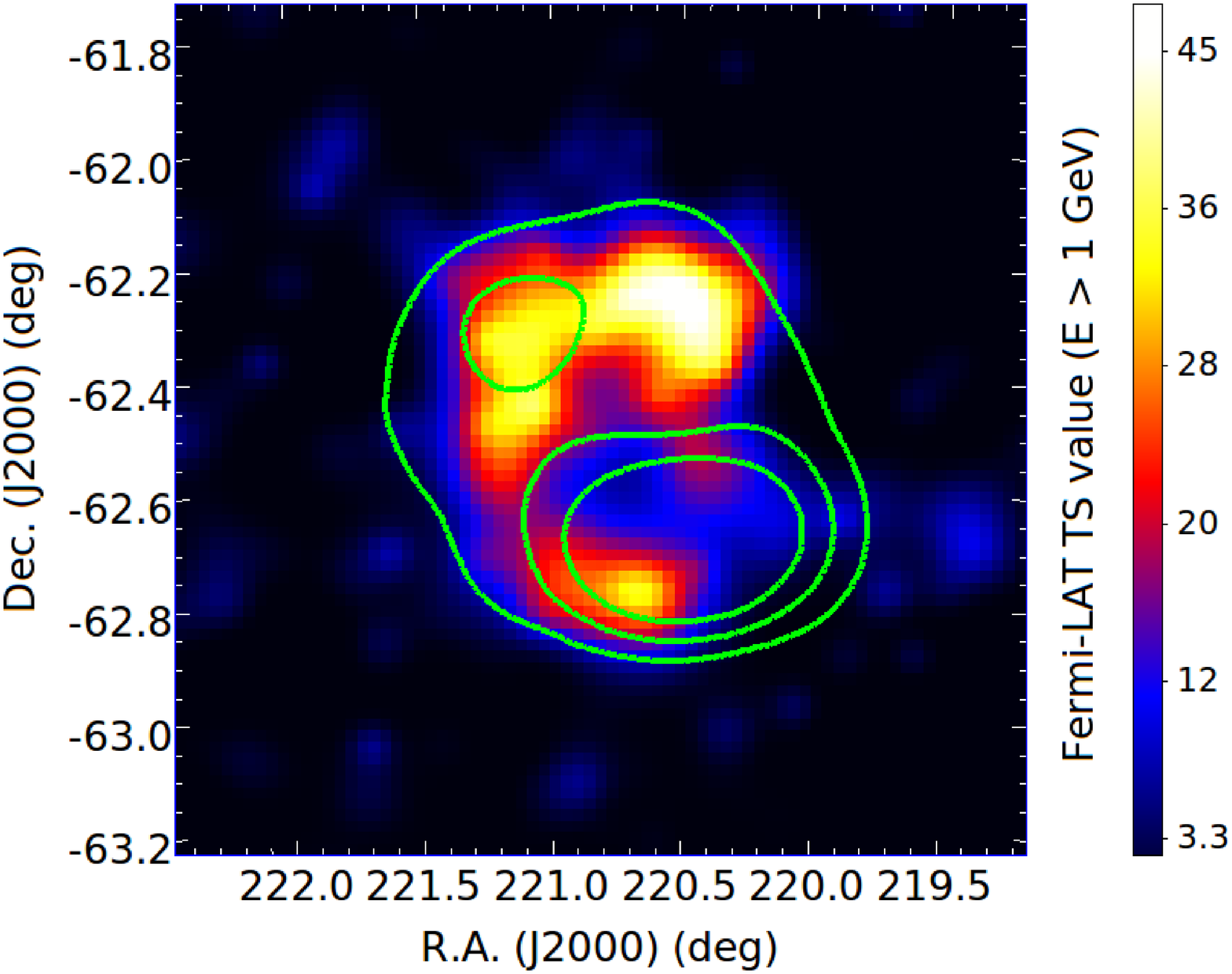}\\
  \end{tabular}
  \caption{Test Statistic ($TS$) map above 1 GeV centered on RCW 86 with MWL contours. All data sets have been smoothed such that their angular resolution is similar to the Fermi-LAT PSF (0$\fdg$27 at 68\% C.L.). The radio (top-left), TeV (top-right), thermal (bottom-left) and non-thermal (bottom-right) X-ray data are from \cite{Murphy2007}, \cite{Abramowski2015} and \cite{Broersen2014} respectively. \label{fig:tsmap_mwl}}
\end{figure}

In addition to geometrical models, we have fitted the LAT data with MWL morphological templates to evaluate the correspondance of the $\gamma$-ray emission above 1 GeV coming from RCW 86 with different source morphologies (see Figure~\ref{fig:tsmap_mwl}). For that purpose, we compared the $TS$ obtained with the best-fit uniform disk model (see Table \ref{tab:results_ptlike}) and the ones obtained with the MWL morphological templates. The radio data are from the Molonglo Observatory Synthesis Telescope (MOST) at 843 MHz \citep{Murphy2007} and the TeV data from H.E.S.S. \citep{Abramowski2015}. Concerning the X-ray data, we used the observations of the space telescope \textit{XMM-Newton}, in the energy range $0.5-1$ keV and $2-5$ keV \citep{Broersen2014}, to estimate the correlation with thermal and non-thermal X-ray emission separately.  The analysis revealed a good match between the HE $\gamma$-ray emission and the VHE $\gamma$-ray and non-thermal X-ray emissions. However, the radio and thermal X-ray signals do not fit well with the HE $\gamma$-ray emission.

%\begin{deluxetable}{lcc}
%\tablecaption{$TS$ values obtained with gtlike above 1 GeV. Numbers in parentheses correspond to the values obtained after fitting separately the two hemispheres shown in Figure~\ref{fig:divided_templates}. \label{tab:results_gtlike}}
%\tablewidth{0pt}
%\tablehead{ 
%\colhead{Template} & \colhead{$TS$} & \colhead{N$_{\rm dof}$}
%}
%\startdata
%    Point Source & 34 & 4 \\
%    Disk  & 97 (99) & 5 (8)\\
%    Ring  & 104 (105) & 6 (9)\\
%    MOST (843 MHz) & 77 (86) & 2 (5)\\
%    \textit{XMM-Newton} ($0.5-1$ keV) & 62 (90) & 2 (5)\\
%    \textit{XMM-Newton} ($2-5$ keV) & 91 (101) & 2 (5)\\
%    H.E.S.S. & 98 (100) & 2 (5)\\
%\enddata
%\end{deluxetable}

\begin{table}[ht]
  \caption{$TS$ values obtained with gtlike above 1 GeV. Numbers in parentheses correspond to the values obtained after fitting separately the two hemispheres shown in Figure~\ref{fig:divided_templates}. \label{tab:results_gtlike}}
  \centering
  \begin{tabular}{lcc}
    \hline
    \hline
    Template & $TS$ & N$_{\rm dof}$\rule[-4pt]{0pt}{12pt}\\
    \hline
    Point Source & 34 & 4\rule[-3pt]{0pt}{12pt}\\
    Disk  & 97 (99) & 5 (8)\rule[-3pt]{0pt}{12pt}\\
    Ring  & 104 (105) & 6 (9)\rule[-3pt]{0pt}{12pt}\\
    MOST (843 MHz) & 77 (86) & 2 (5)\rule[-3pt]{0pt}{12pt}\\
    \textit{XMM-Newton} ($0.5-1$ keV) & 62 (90) & 2 (5)\rule[-3pt]{0pt}{12pt}\\
    \textit{XMM-Newton} ($2-5$ keV) & 91 (101) & 2 (5)\rule[-3pt]{0pt}{12pt}\\
    H.E.S.S. & 98 (100) & 2 (5)\rule[-3pt]{0pt}{12pt}\\
    \hline
  \end{tabular}
\end{table}

\begin{figure}[ht]
  \centering
  \includegraphics[width=\columnwidth]{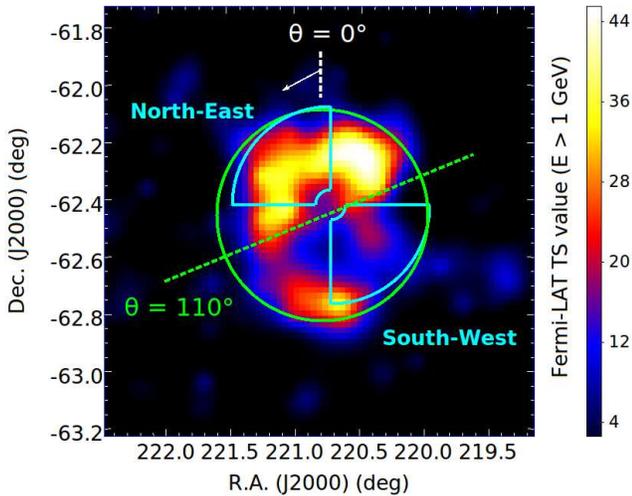}
  \caption{The white dashed line corresponds to the north/south line in equatorial coordinates ($\theta = 0^\circ$) while the green one represents the best angle of division ($\theta = 110^\circ$). The green circle corresponds to the best-fit disk model provided by pointlike in Table \ref{tab:results_ptlike} and the cyan quadrants represent the two regions defined in \cite{Abramowski2015} that are studied in Section 3.2. \label{fig:divided_templates}}
\end{figure}

As RCW 86 is known to present an asymmetry in its morphology from radio to VHE $\gamma$-rays \citep{Broersen2014}, we divided the spatial models and fit separately the two half-templates in order to quantify the difference between the NE and the SW region of the SNR, using the improved PSF of the Pass 8 data. We determined the best (most significant) angle of division by computing the log-likelihood for 18 regularly spaced angles, from $0^\circ$ to $170^\circ$, $0^\circ$ corresponding to a division along the line north/south in equatorial coordinates (the white dashed line in Figure~\ref{fig:divided_templates}). This analysis was performed for all the templates and we obtained the same best angle (the green dashed line in Figure~\ref{fig:divided_templates}) for all of them. 

By comparing the results for non-divided and divided templates presented in Table~\ref{tab:results_gtlike}, we notice that the division improves significantly the likelihood for the MOST and the \textit{XMM-Newton} templates. Moreover, the non-thermal X-ray template is as good as the disk and the H.E.S.S. templates when it is divided. This indicates that the X-ray and radio morphologies do not reproduce well the HE $\gamma$-ray signal in the case of a single region, as it will be confirmed in Section 7 with the broadband modeling of the spectrum. However, the likelihood is not much improved when dividing the disk and the H.E.S.S. template, showing that they well reproduce the whole SNR as seen with \textit{Fermi}-LAT. The non-divided H.E.S.S. template provides the highest TS of all the non-divided (and divided when taking into account the number of degrees of freedom) templates when fitting the HE $\gamma$-ray emission. As a consequence, this template was used to perform the spectral analysis.

\subsection{Spectral analysis}

To study the spectrum of RCW 86, we performed a maximum likelihood fit with \verb+gtlike+ in the energy range 100 MeV -- 500 GeV, using the non-divided H.E.S.S. template as a spatial model. The \textit{Fermi}-LAT data are well described by a power law function ($TS$ = 99), with a photon index of $1.42\ \pm\ 0.1_{\rm stat}\ \pm\ 0.06_{\rm syst}$ and an energy flux above 100 MeV of ($2.91$ $\pm$ $0.8_{\rm stat}$ $\pm$ $0.12_{\rm syst}$) $\times$ $10^{-11}$ erg cm$^{-2}$ s$^{-1}$. One should note that the photon index $\Gamma$ is linked to the radio energy index $\alpha$ (defined as $S_{\nu} \propto \nu^{-\alpha}$) by the relation $\Gamma$ = 1 + $\alpha$. For RCW 86, $\alpha$ is found to be 0.6 and is therefore consistent with a hard index in GeV. We also performed several fits while fixing the index of the power law at different values: 1.5, 1.6, 1.7 and 1.8. In each case, we measured the deterioration of the log-likelihood by computing the difference between the $TS$ of the best-fit and the $TS$ obtained with the fixed index. As a result, we excluded $\Gamma > 1.7$ at more than 3$\sigma$. Although a broken power-law seems more likely when considering both \textit{Fermi}-LAT and H.E.S.S. data, the log-likelihood is improved by only 2$\sigma$ ($\Delta TS$ = 7 for 2 more degrees of freedom) when fitting the spectra with this function. The four additional background sources (Src1, Src2, Src3, and Src4) were taken into account in these fits and their best-fit positions and spectral parameters are given in Table~\ref{tab:specAna_bkgdsrc}. Despite having a $TS$ lower than 25, Src1 and Src2 were kept in our background model to avoid any contamination to the RCW 86 spectrum.

%\begin{deluxetable}{lccccc}
%\tablecaption{Best spectral fit parameters obtained for the nearby background sources Src1, Src2, Src3 and Src4 with gtlike using the H.E.S.S. template for RCW 86 above 100 MeV. The errors are statistical errors only. \label{tab:specAna_bkgdsrc}}
%\tablewidth{\columnwidth}
%\tablehead{ 
%\colhead{Source Name} & \colhead{R.A. ($^\circ$)} & \colhead{Dec. ($^\circ$)} & \colhead{Spectral Index} & \colhead{Flux ($\times$10$^{-12}$ erg/cm$^{2}$/s)} & \colhead{$TS$}
%}
%\startdata
%    Src1 & 220.66 $\pm$ 0.05 & $-$63.21 $\pm$ 0.02 & $1.69 \pm 0.20$ & $5.11^{+3.41}_{-2.51}$ & 24\\
%    Src2 & 223.15 $\pm$ 0.07 & $-$63.00 $\pm$ 0.04 & $1.82 \pm 0.24$ & $4.19^{+3.41}_{-2.25}$ & 18\\
%    Src3 & 224.15 $\pm$ 0.14 & $-$63.22 $\pm$ 0.05 & $2.56 \pm 0.13$ & $7.11^{+1.53}_{-1.65}$ & 26\\
%    Src4 & 222.50 $\pm$ 0.09 & $-$60.36 $\pm$ 0.05 & $2.54 \pm 0.12$ & $15.3^{+2.67}_{-2.61}$ & 52\\
%\enddata
%\end{deluxetable}

\begin{table*}[ht]
  \caption{Best spectral fit parameters obtained for the nearby background sources Src1, Src2, Src3 and Src4 with gtlike using the H.E.S.S. template for RCW 86 above 100 MeV. The errors are statistical errors only. \label{tab:specAna_bkgdsrc}}
  \centering
  \begin{tabular}{lccccc}
    \hline
    \hline
    %\colhead{Source Name} & \colhead{R.A. ($^\circ$)} & \colhead{Dec. ($^\circ$)} & \colhead{Spectral Index} & \colhead{Flux ($\times$10$^{-12}$ erg/cm$^{2}$/s)} & \colhead{$TS$}
    Source Name & R.A. ($^\circ$)& Dec. ($^\circ$)& Spectral Index & Flux ($\times$10$^{-12}$ erg/cm$^{2}$/s) & $TS$\rule[-4pt]{0pt}{12pt}\\
    \hline
    Src1 & 220.66 $\pm$ 0.05 & $-$63.21 $\pm$ 0.02 & $1.69 \pm 0.20$ & $5.11^{+3.41}_{-2.51}$ & 24\rule[-3pt]{0pt}{12pt}\\
    Src2 & 223.15 $\pm$ 0.07 & $-$63.00 $\pm$ 0.04 & $1.82 \pm 0.24$ & $4.19^{+3.41}_{-2.25}$ & 18\rule[-3pt]{0pt}{12pt}\\
    Src3 & 224.15 $\pm$ 0.14 & $-$63.22 $\pm$ 0.05 & $2.56 \pm 0.13$ & $7.11^{+1.53}_{-1.65}$ & 26\rule[-3pt]{0pt}{12pt}\\
    Src4 & 222.50 $\pm$ 0.09 & $-$60.36 $\pm$ 0.05 & $2.54 \pm 0.12$ & $15.3^{+2.67}_{-2.61}$ & 52\rule[-3pt]{0pt}{12pt}\\
    \hline
  \end{tabular}
\end{table*}

Systematic errors are defined as Err$_{\rm syst} = \sqrt{(Err_{\rm iem})^2 + (Err_{\rm irf})^2 + (Err_{\rm model})^2}$. This expression takes into account the imperfection of the Galactic diffuse emission model (Err$_{\rm iem}$), uncertainties in the effective area calibration (Err$_{\rm irf}$) and uncertainties on the source shape (Err$_{\rm model}$). The first one was estimated by using alternative Interstellar Emission Models (IEM), as described in \cite{SNRCAT2015}. For the second one, we applied scaling functions that change the effective area \citep[$\pm$ 3\% for 100 MeV -- 100 GeV and $\pm$ $3\% + 10\% \times (\log (E/{\rm MeV})-5)$ above 100 GeV][]{Fermi2012}. The third one was obtained by fitting the $\gamma$-ray emission with the best spatial models (the disk and the ring) provided by \verb+pointlike+.

\begin{figure}[ht]
  \centering
  \includegraphics[width=\columnwidth]{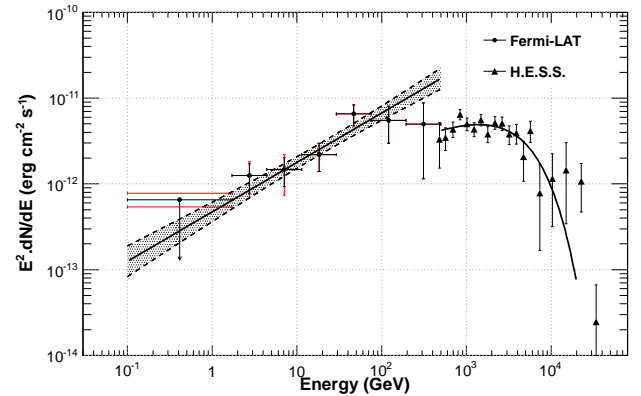}
  \caption{Spectral Energy Distribution of RCW 86 from 100 MeV to 50 TeV with the Fermi-LAT and H.E.S.S. \citep{Abramowski2015} data points shown as black circles and triangles, respectively. The black solid line passing through the H.E.S.S. points corresponds to the best-fit spectrum from \cite{Abramowski2015}. The smaller and larger errors on the \textit{Fermi} fluxes are statistical and total errors (quadratic sum of statistical and systematic errors), respectively. The range of upper limit values (black and red lines) correspond to the uncertainty in the diffuse modeling. The dark gray shaded area (delimited by black dashed lines) represents the 68\% confidence band of the fitted \textit{Fermi}-LAT spectrum.\label{fig:sed}}
\end{figure}

Figure~\ref{fig:sed} shows the Spectral Energy Distribution (SED) of RCW 86. The \textit{Fermi}-LAT spectral points were obtained by dividing the 100 MeV -- 500 GeV range into nine logarithmically spaced energy bins. We assumed a power law with a spectral index of 1.4 and used the H.E.S.S. spatial model. In addition to RCW 86, only very bright and close sources ($TS >$ 500 and within a radius of 5$^\circ$), as well as the Galactic diffuse and the isotropic emission were fitted. We derived the 95\% C.L. upper limit in the 0.1 -- 1.7 GeV band, combining the first three bins in which no signal was detected by the \textit{Fermi}-LAT.

To pursue our purpose of understanding the variation of the physical conditions in RCW 86, we performed a spectral analysis between 100 MeV and 500 GeV using the H.E.S.S. template divided in half along the green dashed line shown in Figure~\ref{fig:divided_templates}. We obtained an index of 1.36 $\pm$ 0.17$_{\rm stat}$ and an energy flux above 100 MeV of (1.69 $\pm$ 0.63) $\times$ 10$^{-11}$ erg cm$^{-2}$ s$^{-1}$ for the upper region and an index of 1.62 $\pm$ 0.17$_{\rm stat}$ and an energy flux of (0.97 $\pm$ 0.31) $\times$ 10$^{-11}$ erg cm$^{-2}$ s$^{-1}$ for the lower region. The index seems to be harder in the upper region but there is no significant difference when taking into account the errors. In addition to that, we also studied the spectrum of the two specific areas that were defined in \cite{Abramowski2015}, as shown in Figure~\ref{fig:divided_templates}. The NE and SW quadrants were fitted in two different fits. For each quadrant we subtracted it from the rest of the disk and fitted both the quadrant and the complementary region simultaneously. The analysis of the \textit{Fermi}-LAT data revealed significant $\gamma$-ray emission at $\sim 4.7 \sigma$ ($TS$ = 22) in the NE region but no signal was detected in the SW region ($TS <$ 3). The spectrum of the NE signal is well-fitted by a power law function with a hard index of $1.33 \pm 0.20 $ and an energy flux of $(1.2 \pm 0.5_{\rm stat})$ $\times$ 10$^{-11}$ erg cm$^{-2}$ s$^{-1}$ and we derived a 95\% C.L. upper limit for the SW region (1.09 $\times$ 10$^{-12}$ erg cm$^{-2}$ s$^{-1}$).

%%%========================================================================================================%%%
\section{Radio Continuum Data}

RCW~86 is included in the second Molonglo Galactic Plane Survey (called hereafter MGPS-2) performed by the Molonglo Observatory Synthesis Telescope (MOST), at 843~MHz with a bandwidth of 3 MHz and a resolution of $\approx 45 \times 51$~arcsec$^2$ \citep{Murphy2007}. However, the MOST data are missing structures on scales larger than $\sim 20{-}30$~arcmin, and this survey does not recover the total radio emission from RCW~86 \cite[its integrated flux density is $\approx 20$~Jy in MGPS-2, whereas $\approx 55$~Jy is expected from previous observations, e.g.][]{Caswell1975}. Instead, to obtain radio flux densities for regions of the SNR, we used Parkes survey observations at 2.4~GHz, with a resolution of $10.2 \times 10.6$~arcmin$^2$, from \cite{Duncan1995}. The integrated flux density of RCW~86 in this survey is $\approx 25$~Jy, in reasonable agreement with that expected, showing that this single-dish survey is not missing flux from this source. Given the low resolution of this survey, the flux densities in the NE and SW quadrants of \cite{Abramowski2015} were obtained by integrating out to somewhat larger radii of 40~arcmin, which gives 4.3~Jy in the NE quadrant and 10.4~Jy in the SW quadrant.

%%%========================================================================================================%%%
\section{X-ray observations}

\begin{figure}[ht]
  \centering
  \includegraphics[height=2.4in]{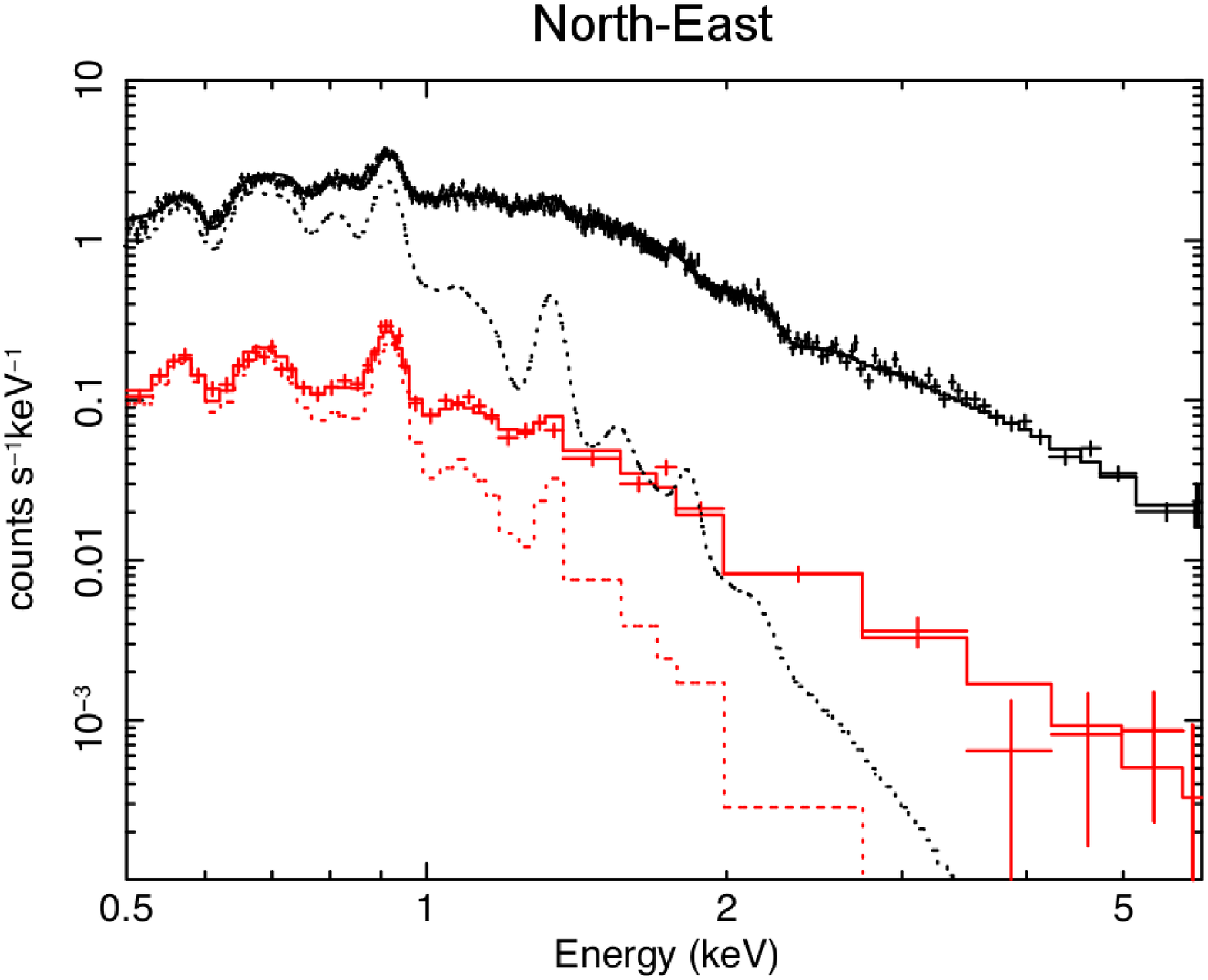}
  \includegraphics[height=2.4in]{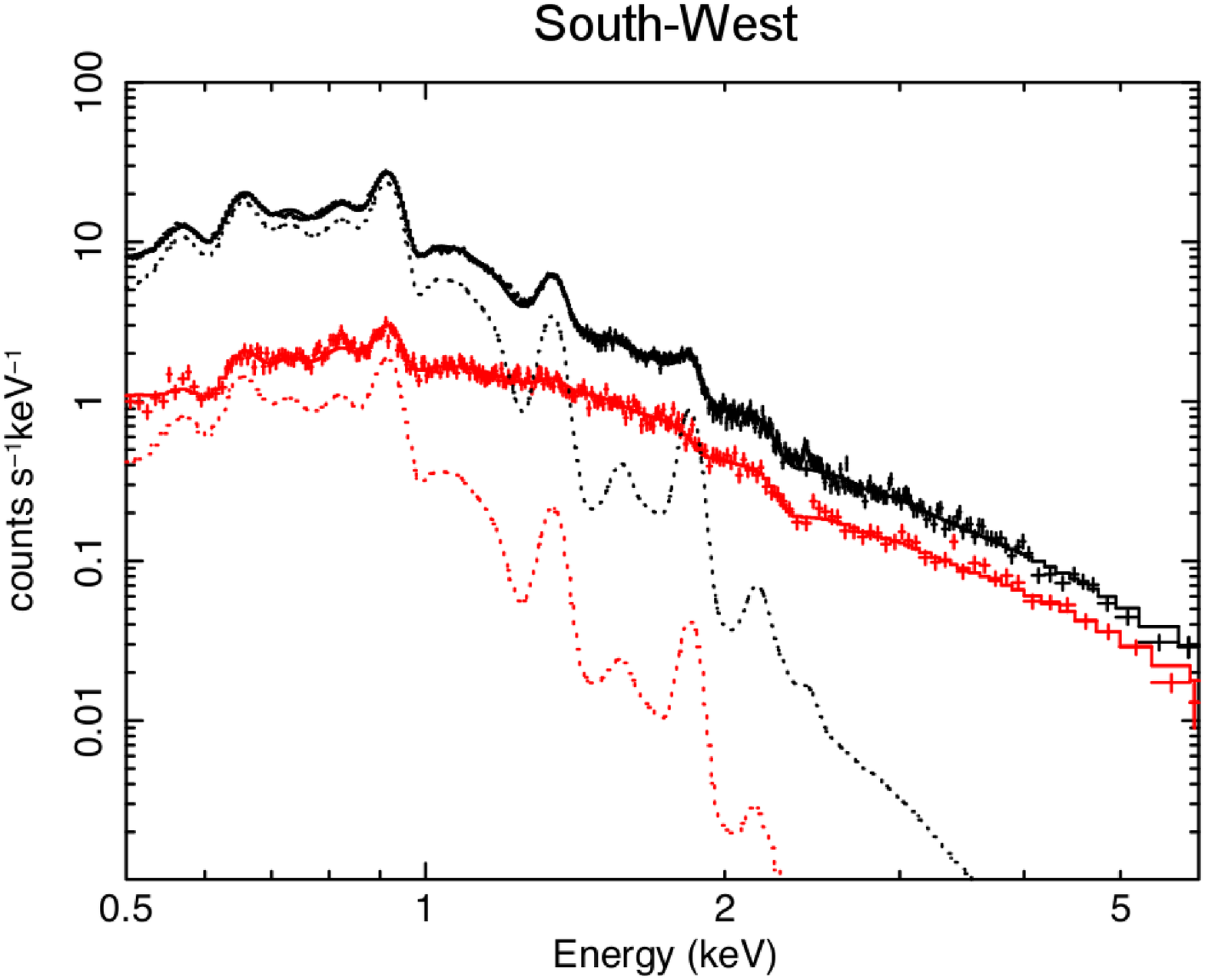}
  \caption{Observed spectra for the split northern regions (Top) and the southern regions (Bottom). Due to the limited field of view of \textit{XMM-Newton}, two pointings were needed to cover each quadrant shown in Figure~\ref{fig:divided_templates}, hence the black and red curves in both plots. Solid lines show the spectral fits while the dotted lines give the contributions of the thermal components, which are negligible above 2 keV.\label{fig:plot_xrays}}
\end{figure}

To estimate the non-thermal X-ray emission from the NE and SW regions of RCW 86, we analyzed the spectra of these two regions using data of the EPIC-MOS2 instrument of XMM-Newton. Since the SNR is larger than the XMM-Newton field of view the NE and SW regions were split over several observations, and we took care of this by using spectral analysis of mutually exclusive regions, which together overlapped entirely with the regions indicated in Figure~\ref{fig:divided_templates}. For the NE  we used  observations number 0208000101 (Jan. 26th 2004, 59.992 ks) and  0504810301 (Aug. 25th 2007, 72.762 ks). For the SW region we used observations 0110010701 (Aug. 16th 2000, 23.314 ks) and 0504810401 (Aug. 23rd 2007, 116.782 ks). The spectra presented in Figure~\ref{fig:plot_xrays} were extracted using the standard XMM-Newton analysis package XMM SAS, version 14.0 \footnote{See http://xmm.esac.esa.int/sas/ for more information about the software.}. The background spectra were obtained from empty regions in the field, taken from the same observations. The extracted spectra were then analyzed  with the spectral analysis software xspec 12.8 \citep{Arnaud1996} using the ``vnei'' model plus power law component, both corrected for Galactic absorption. The absorption columns are (2.5 $\pm$ 0.1)$\times$ 10$^{21}$ cm$^{-2}$ for the southern part and (4.8 $\pm$ 0.1) $\times$ 10$^{21}$ cm$^{-2}$ for the northern part. From the best fit models we obtained the fluxes in the 3-5 keV band, which for RCW 86 is totally dominated by SC emission. The fluxes are estimated to be $10.0 \times 10^{-12}$ erg cm$^{-2}$ s$^{-1}$ and $4.3 \times 10^{-12}$ erg cm$^{-2}$ s$^{-1}$ for the SW and NE regions, respectively. The flux measurements have errors of the order of 5-10\%, mostly dominated by systematic errors, as absolute flux calibration of X-ray instruments is accurate at the 5\% level. For both regions, the power law index is measured to be $3.0 \pm 0.2$, with the error mostly due to variations within each region.

%%%========================================================================================================%%%
\section{The surrounding interstellar medium} 

We have analyzed the cold neutral gas in the environs of RCW 86 to investigate the characteristics of the surrounding gas. To carry out this search we used data at $\lambda$ = 21 cm acquired with the Australia Telescope Compact Array (ATCA) on March 24, 2002. To recover the missing short spatial frequencies, the ATCA data were combined in the u-v plane with single dish observations performed with the Parkes radio telescope. The final data are arranged in a cube with an angular resolution of $2^\prime.7 \times 2^\prime.5$ (R.A. $\times$ Dec.) and a 1$\sigma$ rms noise in line-free channels of about 1 K. The cube covers the velocity range -120.00 to +126.00 km s$^{-1}$ with a velocity resolution of 0.82 km s$^{-1}$. 

The whole H{\sc i} cube was inspected searching for imprints in the surrounding medium that might have been produced by the SN explosion and/or its precursor star. Different morphological signatures can be left in the interstellar gas by these expanding events, like cavities blown up by the stellar wind of the pre-supernova, bubbles surrounded by a higher density neutral shell, accelerated clouds seen in projection against the center of the SNR, etc. These kind of features have been identified  in association with several Galactic SNRs (see e.g. \cite{Park2013} and references therein). In the case of RCW 86 we detected the presence of an elongated cavity, about 1$\fdg$5 in size, that runs almost parallel to the Galactic plane, in the velocity interval between $\sim -38$ km~s$^{-1}$ and $\sim -32$ km~s$^{-1}$ (all velocities are referred to the Local Standard of Rest, LSR). Within this velocity range, more precisely between $\sim -35$ km~s$^{-1}$ and $\sim -33$ km~s$^{-1}$  the SNR appears surrounded by a tenuous, approximately circular H{\sc i} shell with variable brightness distribution. These morphological findings are in very good agreement with the predictions made on the basis of radio continuum, X-rays, infrared observations and hydrodynamic simulations \citep{Vink1997, Dickel2001, Williams2011, Broersen2014}. After applying a circular rotation model for our Galaxy for $l = 315\fdg4$, $b = -2\fdg3$, the LSR radial velocity interval of the observed features translates into a distance of $\sim 2.5 \pm 0.3$ kpc. This distance is in very good agreement with that previously obtained for RCW 86 on the basis of optical measurements of proper motions of the filaments \citep{Rosado1996, Sollerman2003}, suggesting that this gas is placed at the same distance as RCW 86. In addition to the morphological signatures, an independent test of the adopted central radial velocity can be done by comparing the absorbing column density N$_{\rm H}$ integrated between us and the SNR with Yamaguchi's (2011) best fits derived from X-ray observations. From our H{\sc i} data we obtain N$_{\rm H}$ = 2.6 $\times 10^{21}$ cm$^{-2}$ (for the whole annulus shown in Figure~\ref{fig:annulus}), in good concordance with the values N$_{\rm H}$ = 2.9 or 2.8 $\pm$ 0.3 $\times 10^{21}$ cm$^{-2}$, obtained from X-ray data (where the different values depend on the model) and the absorption column given in Section 5 for the southern region. The apparent discrepancy with the absorption columns mentioned in Section 5 for the northern region can be due to the presence of H2, a contribution to which $\lambda$ 21 cm observations are not sensitive. As the distribution of the molecular gas (CO in particular) perpendicular to the plane has a semi-scale of $\sim$ 55 pc in the inner Galaxy, the presence of some molecular gas fragmented in small isolated clouds is natural at the height of RCW 86 ($\sim$ 100 pc). The line of sight where the X-ray absorption was calculated might have crossed one of these cloudlets. Overall, the fact that these H{\sc i} features simultaneously fulfill morphological and kinematical criteria, strongly suggest that the neutral gas observed in this velocity interval is physically associated with the SNR.

\begin{figure}[ht]
  \centering
  \includegraphics[width=\columnwidth]{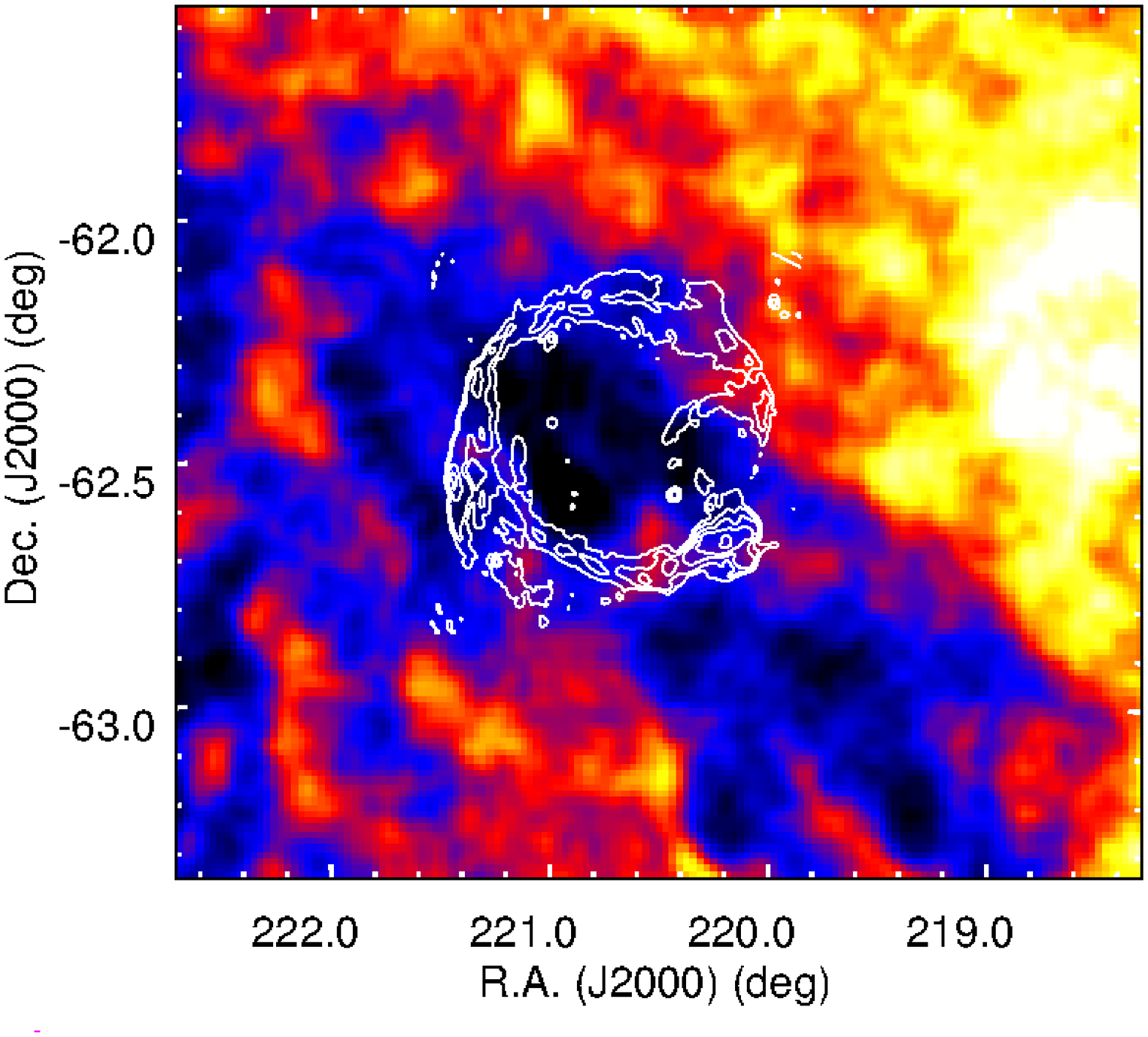}
  \includegraphics[width=\columnwidth]{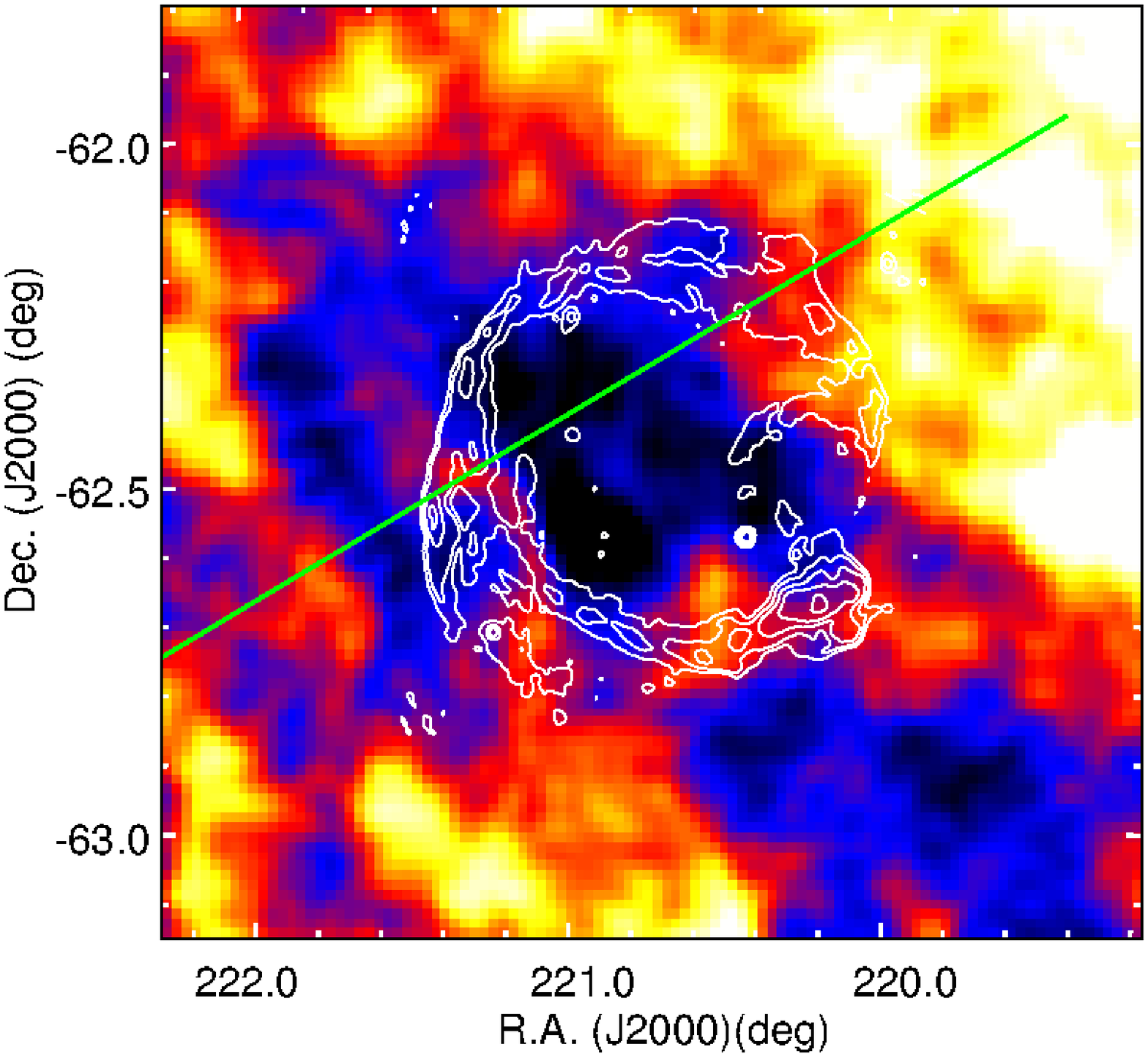}
  \caption{Neutral hydrogen distribution around RCW 86. Top: the HI distribution is displayed with a linear scale between 32.8 and 89.8 K in an ample field around RCW 86 at the LSR radial velocity of -34  km s$^{-1}$. Bottom: the same, but in a smaller field in the vicinity of RCW 86, displayed between 32.8 and 73.3 K to emphasize the presence of the internal shell. The white contours show the radio continuum emission at 843 MHz from the MGPS-2 data. The line included in the bottom panel indicates the direction where the distribution profile shown in Figure~\ref{fig:radial_profile} was extracted. \label{fig:nhd}}
\end{figure}

Figure~\ref{fig:nhd} (Top) shows the local H{\sc i} distribution in a large field (over 5 square degrees) in direction to RCW 86, as observed around the  radial velocity of $\sim - 34$ km s$^{-1}$. The white contours show the radio continuum emission at 843 MHz from the MGPS-2 data.  Figure~\ref{fig:nhd} (Bottom) shows the same as in Fig.~\ref{fig:nhd} (Top) but in a smaller region around RCW 86 and using a different scale so as to emphasize the fainter inner shell. Figure 7 displays a  radial profile traced across the line shown in Fig.~\ref{fig:nhd} (Bottom). The arrows indicate the approximate locations of the walls of the outer cavity and  the inner shell.
 
The H{\sc i} observations can be used to carry out independent estimates of the volume density of the SNR environs. We considered four regions corresponding to the four quadrants of two concentric circles traced with inner and  outer radii coincident with the radio shell (as shown in Figure~\ref{fig:annulus}). In this estimate, two aspects have to be considered: the background emission contribution, that takes into account emission that may come from far gas whose emission is detected at the same radial velocities in this direction of the Galaxy, and the geometry of the associated gas along the line of sight. For the first issue we subtracted a uniform background of T=25 K, a value estimated from the inspection in the observed field of regions free of structures down to the angular resolution of the data. This assumption is reasonable since the neutral gas located at the far distance is at a height well below the Galactic plane in the direction to $b=-2\fdg3$. Concerning the three-dimensional distribution of the adjacent gas, we tested the two usually adopted geometries: a cylindrical ring with a depth similar to the SNR diameter (the case where the gas accompanies a barrel-shaped SNR), and a spherical shell surrounding the SNR. This last case is consistent with the geometry suggested by the Balmer-dominated filaments that encircle almost the complete periphery of RCW86 \citep{Smith1997}. It is known that the Balmer-dominated filaments arise from relatively high velocity shocks passing through partially neutral gas. It is natural then to assume that the cold H{\sc i} mimics the optically depicted SNR. In any case in both considered geometries the results obtained were very similar and in what follows we list an average between the two results. 

For region 1 (NW quadrant) n$_{\rm H} \sim 1.5$ cm$^{-3}$; for region 2 (NE quadrant) n$_{\rm H} \sim 1$ cm$^{-3}$; for region 3 (SE quadrant) n$_{\rm H} \sim 1$ cm$^{-3}$ and for region 4 (SW quadrant) n$_{\rm H} \sim 1.2$ cm$^{-3}$. In all cases the intrinsic error of the quoted numbers is of about 30\% taking into account the uncertainty in the distance (of 25\%) and the approximate background subtraction. For the interior of the SNR  we estimated  n$_{\rm H} \sim 0.5 $ cm$^{-3}$.

The complete analysis of the H{\sc i} in the direction of RCW 86 will be published elsewhere.

\begin{figure}[ht]
  \centering
  \includegraphics[width=0.9\columnwidth]{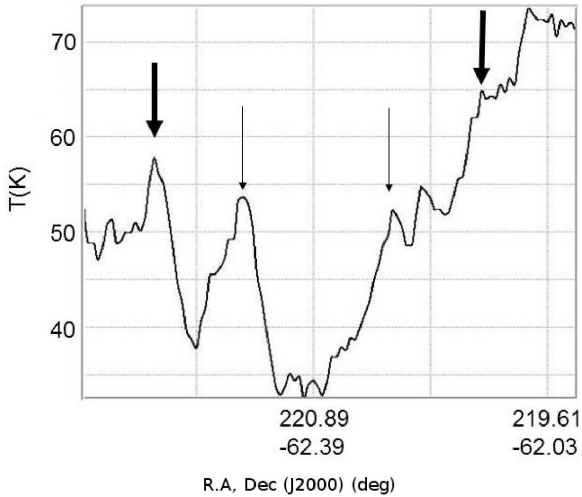}
  \caption{Radial profile of the HI distribution measured along the green line showed in Figure~\ref{fig:nhd} (Right). The arrows indicate the presence of an inner shell within a more extended cavity. The x-axis corresponds to the equatorial coordinates (in degrees) while the y-axis shows the temperature (in kelvin). \label{fig:radial_profile}} 
\end{figure}

\begin{figure}[ht]
  \centering
  \includegraphics[width=0.9\columnwidth]{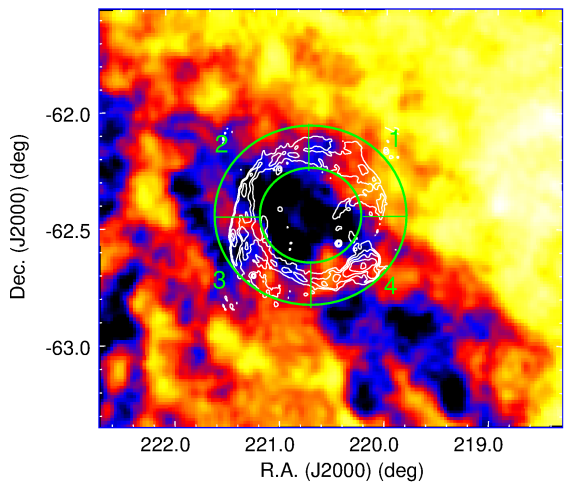}
  \caption{Neutral hydrogen distribution around RCW 86 at the LSR radial velocity of -34 km s$^{-1}$. The green annulus delimits the regions where the atomic density was estimated.\label{fig:annulus}}
\end{figure}

%%%========================================================================================================%%%
\section{Discussion}

\subsection{Broadband modeling}

The main difficulty in determining the origin of the $\gamma$-ray emission of RCW 86 lies in the competition of two major channels of $\gamma$-ray production: the IC scattering of high energy leptons on local photon fields (leptonic scenario) or the decay of neutral pions produced by the interaction between accelerated protons and interstellar clouds located near the remnant (hadronic scenario). We performed a broadband modeling of the non-thermal emission of RCW 86 using radio, X-ray and VHE $\gamma$-ray data, in addition to the \textit{Fermi}-LAT observations. This modeling (shown in Figure~\ref{fig:sed_modeling}) aims to constrain key parameters such as the average magnetic field and the fraction of the total SN energy which is transferred into protons and electrons. The particle spectra are assumed to follow a power-law function with an exponential cutoff dN/dE $\propto$ E$^{-\Gamma} \times$ exp(-E/E$_{\rm max}$) with the same index for both distributions (electron and proton), starting at 511 keV for the electrons and 1 GeV for protons. The escape of accelerated particles confined in the magnetic field of the shock is taken into account, assuming a shell thickness of $0\fdg1$. This value was obtained by fitting the \textit{Fermi}-LAT data with a ring model and is in agreement with Abramowski et al.'s 2015 estimates. We define $\eta_{\rm e,n}$ as the ratio of the total energy injected into accelerated particles $W_{\rm e,p}$ to the standard energy of a Type Ia SN explosion $E_{\rm SN}$, assumed to be 10$^{51}$. The so-called electron-to-proton ratio $K_{\rm ep}$ is also computed, at momentum 1 GeV c$^{-1}$ and may be compared to the value measured in cosmic-rays ($K_{\rm ep} \sim$ 10$^{-2}$).

\subsection{Modeling of the whole SNR}

Here we present the results of two leptonic scenario models. Since a pure hadronic scenario requires unlikely parameter values such as a very hard spectral index for protons \citep[as it was already suggested in][]{Lemoine2012} and a high magnetic field ($B > 100$~$\mu$G), we did not consider this case. The presence of a high magnetic field is not excluded in very thin regions, near the shock, but it is very unlikely to have such high values for the whole remnant. Moreover, a spectral index softer than 1.7 is excluded with more than 3$\sigma$, as described in Section 3.2. The hadronic model relying on the interactions between escaped protons and a dense interstellar medium in the vicinity of the remnant, as proposed in \cite{Gabici2009}, seems also ruled out by the non-detection of molecular clouds in the NE part of the remnant, where the $\gamma$-ray signal detected by the \textit{Fermi}-LAT is the most important. 

\begin{figure}[ht]
  \centering
  \includegraphics[height=3.2in]{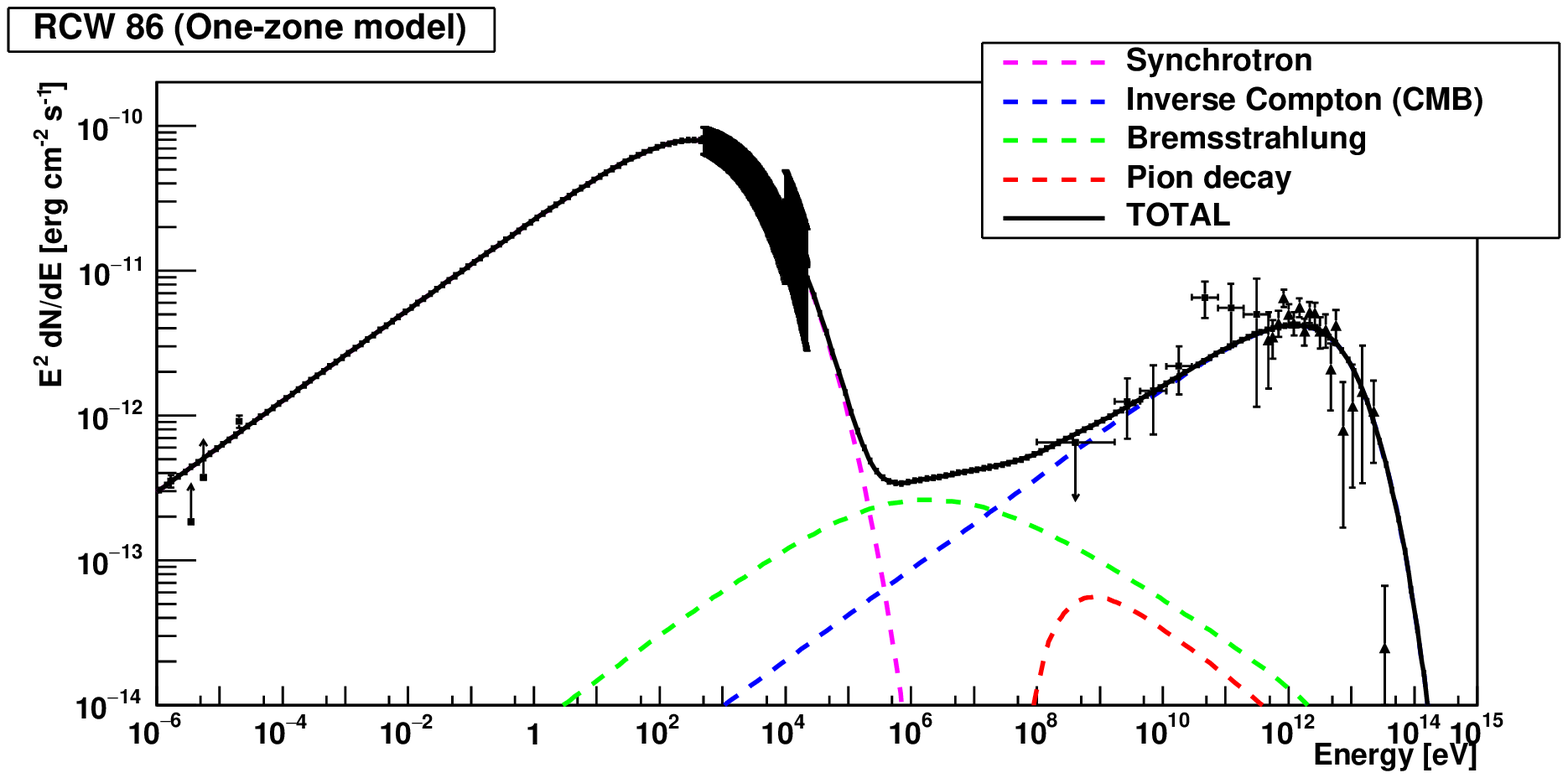}\\
  \includegraphics[height=3.2in]{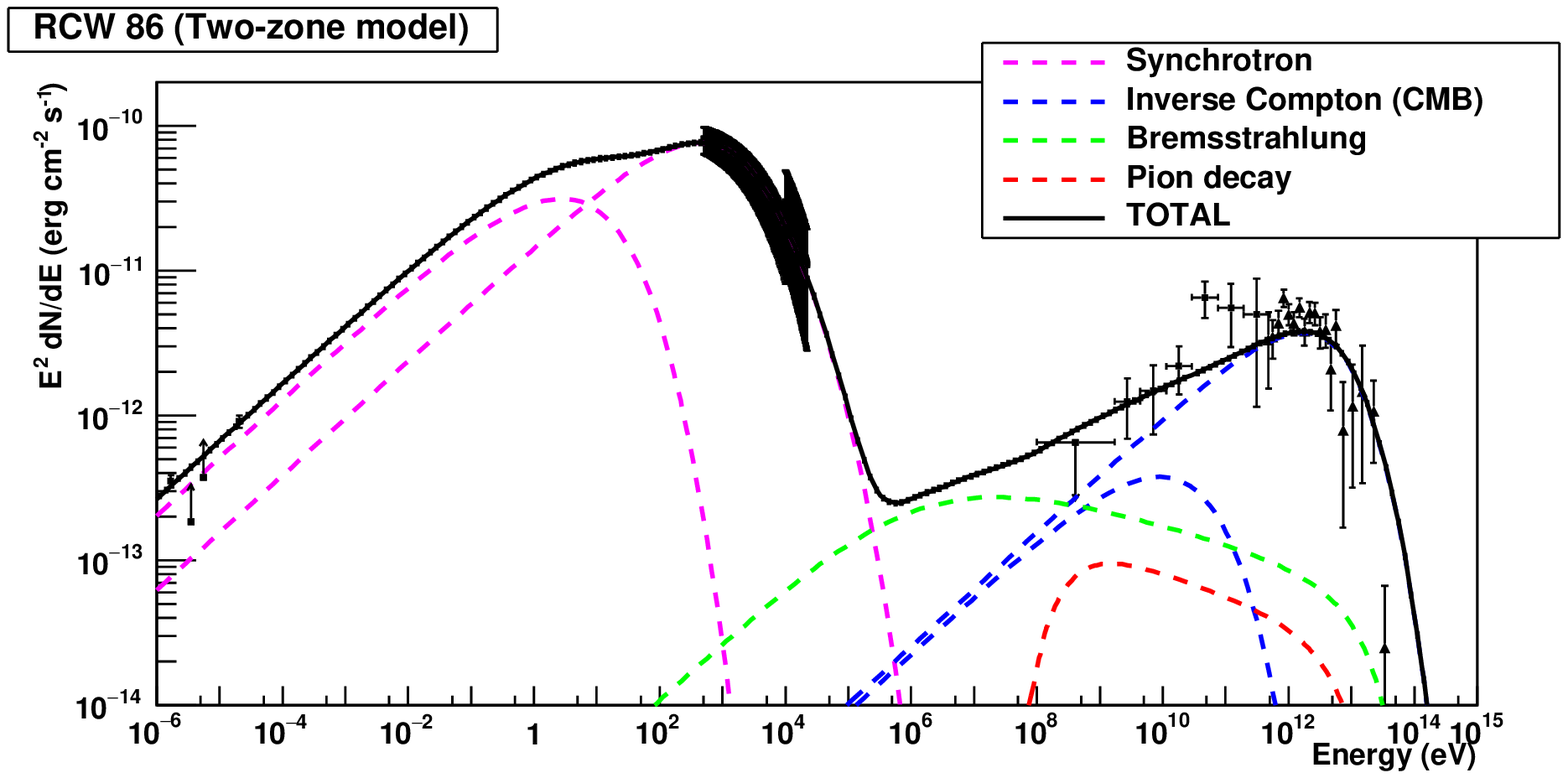}
  \caption{Spectral Energy Distribution of RCW 86 with the best-fit leptonic one-zone (top) and two-zone (bottom) models. The radio points are from Molonglo at 408 MHz and Parkes at 5 GHz \citep{Caswell1975, Lemoine2012} and lower limits from MOST at  843 MHz and ATCA at 1.43 GHz \citep{Whiteoak1996, Dickel2001}. X-ray spectra from ASCA and RXTE are from \cite{Lemoine2012}. The \textit{Fermi}-LAT spectral points and upper limits (95\% C.L.) derived in Section 3.2 and the H.E.S.S. data points in the VHE $\gamma$-ray domain from \cite{Abramowski2015} are also shown. \label{fig:sed_modeling}}
\end{figure}

Figure~\ref{fig:sed_modeling} shows the result for a one-zone model (top) in which we assumed that SC and IC photons are produced by electrons confined in the same emitting region with a constant magnetic field and a two-zone model (bottom) in which we considered two different populations of radio, X-ray and $\gamma$-ray emitting particles. Parameters of the latter population were obtained with a $\chi ^2$ fit without considering the radio emitting population for which parameters were determined afterwards in respect to the previous results. The two-zone model is motivated by the bad correlation between the radio and the $\gamma$-ray data, as shown in Section 3, and by several publications which reported large variations of the physical conditions in RCW 86 \citep{Vink2006, Broersen2014}. To be more conservative on the fraction of the energy injected in protons, the only photon field that was taken into account for the IC scattering of electrons is the Cosmic Microwave Background (CMB). The best-fit parameters for these two models are given in Table~\ref{tab:models_parameters}. For the whole remnant, the radio points in the one-zone model imply a soft spectral index ($\sim$ 2.4) which would lead to a very strong bremsstrahlung component below 1 GeV. To reconcile this low energy component with the new \textit{Fermi}-LAT upper limit at 1 GeV derived in Section 3.2, a maximum density of 0.1 cm$^{-3}$ needs to be assumed. In the case of a two-zone model, we obtained a more reasonable index ($\sim$2.2) with a density of 1.0 cm$^{-3}$ and an energy of 2\% of $E_{\rm SN}$ that goes into protons. Considering a distance of 2.5 kpc and a shell thickness of $0\fdg1$, the energy density of CRs is estimated at $\sim$40 eV cm$^{-3}$. In both cases, the magnetic field is around 10 $\mu$G, in agreement with previous modeling by \cite{Lemoine2012}, \cite{Yuan2014} and \cite{Abramowski2015}.

\subsection{Modeling of the NE and SW regions}

\begin{figure}[ht]
  \centering
  \includegraphics[height=3in]{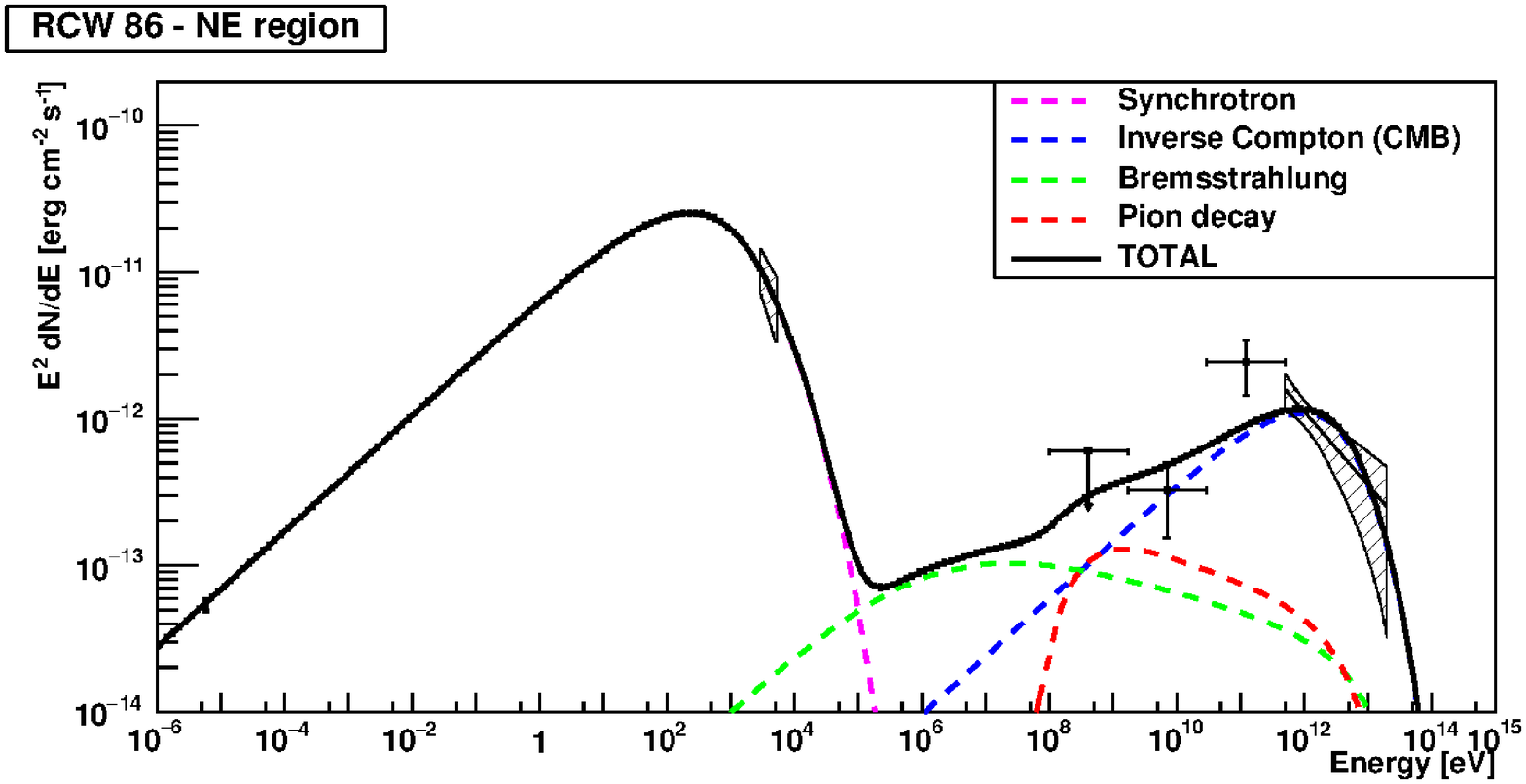}\\
  \includegraphics[height=3in]{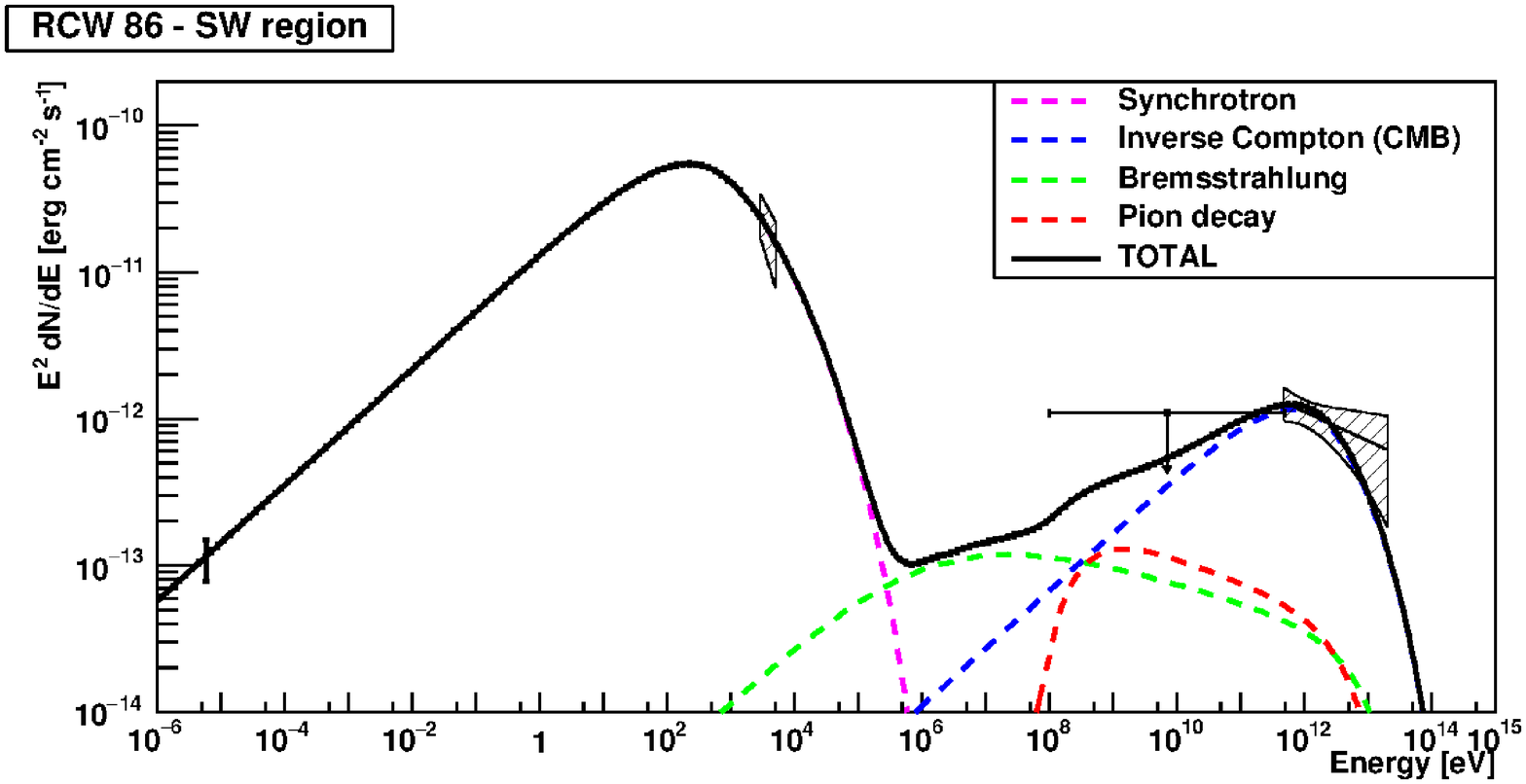}
  \caption{SED modeling for the NE (top) and SW (bottom) regions of RCW 86. The radio fluxes were derived from Parkes observations at 2.4 GHz (as described in Section 4), the X-ray fluxes are estimated using XMM-Newton/MOS data (see Section 5) and the TeV points are from \cite{Abramowski2015}. \label{fig:sed_modeling_regions}}
\end{figure}

In addition to the modeling of the whole SNR, we studied the broadband signal emitted by the NE and SW regions defined in \cite{Abramowski2015} and for which we performed a spectral analysis in Section 3.2 of this paper. In this work, we gathered radio (Section 4), X-ray (Section 5), GeV (Section 3.2) and TeV data \citep{Abramowski2015} for these two regions and performed a modeling assuming that each region sees a different population of emitting particles. The spectral points for the NE region were obtained by dividing the 100 MeV -- 500 GeV energy range into three logarithmically spaced bins only (instead of nine for the whole remnant) because of the reduced statistics. We derived a 95\% C.L upper limit for the first bin and obtained significant fluxes for the two other bins. To limit the number of fitted parameters, the density was assumed to be of 1.0 cm$^{-3}$ for both regions (which is in agreement with the values derived in Section 6). We decided to use the best index previously obtained for the whole remnant, in the case of a two-zone model, (2.21) and fixed it for both electron and proton distributions for the two regions, since there is no evidence in favour of different injection slopes in the remnant. The energy injected in protons was fixed to 0.5\% of $E_{\rm SN}$ (which corresponds to a quarter of the value used for the whole SNR) and the density at 1.0 cm$^{-3}$. Results are shown in Figure~\ref{fig:sed_modeling_regions} and Table~\ref{tab:models_parameters} summarizes the parameters for the two models, obtained with a $\chi ^2$ fit. 

We can notice that the magnetic field is slightly higher in the SW than in the NE, implying a magnetic field gradient in a direction away from the Galactic Plane possibly due to the shock interaction with a denser medium. Another interesting point is that the magnetic field of the NE region is very close to the one obtained for the whole SNR, which is consistent with the fact that most of the GeV emission is detected in the NE part of the remnant. Moreover, $E_{\rm max}$ is also higher in the SW than in the NE which is in agreement with the values of the magnetic field: at early times, when the maximum energy is not limited by SC losses, a higher magnetic field implies a higher $E_{\rm max}$. Overall, the MWL data indicate variations of the magnetic field within the SNR. The radio emission corresponds to regions with high magnetic fields whereas the GeV emission detected by \textit{Fermi}-LAT corresponds to regions with mixed magnetic fields. And since the H.E.S.S. map shows brighter emission coming from the inside of the remnant than in radio and X-rays, the reverse shock could also be responsible for the CR acceleration but with a lower magnetic field. In the near future, a deep study of RCW 86 with the Cherenkov Telescope Array \citep{CTA2011} and ASTRO-H \citep{Takahashi2014} could constrain the magnetic field and provide a precise map of its fluctuation at smaller scales.

%%%========================================================================================================%%%
\section{Conclusions}

Analyzing more than 6 years of \textit{Fermi}-LAT Pass 8 data, we present the first deep study of the morphology and spectrum of the young SNR RCW 86. The spectrum is described by a pure power-law function with an index of $1.42 \pm 0.1_{\rm stat} \pm 0.06_{\rm syst}$ in the LAT energy range (0.1-500 GeV). The broadband emission from radio to TeV cannot be described by a pure hadronic scenario due to the very hard spectral index in the GeV range, the high magnetic field needed and the lack of a high density medium. The two-zone model provides new constraints on the fraction of the total energy injected in protons and the most conservative value amounts to $\sim 2 \times 10^{49}$ erg for a density of 1 cm$^{-3}$. Finally, the non-detection of the SW region of RCW 86, which is very bright in radio, X-rays and at TeV energies, provides specific constraints on this part of the remnant, in terms of the acceleration mechanism as well as the gas density and the magnetic field.
%rises constraints concerning the mechanism occurring in this part of the remnant, as well as about the characteristics of the medium such as the density and the magnetic field.

\textit{Acknowledgments}. The \textit{Fermi} LAT Collaboration acknowledges generous ongoing support from a number of agencies and institutes that have supported both the development and the operation of the LAT as well as scientific data analysis. These include the National Aeronautics and Space Administration and the Department of Energy in the United States, the Commissariat \`a l'Energie Atomique and the Centre National de la Recherche Scientifique / Institut National de Physique Nucl\'eaire et de Physique des Particules in France, the Agenzia Spaziale Italiana and the Istituto Nazionale di Fisica Nucleare in Italy, the Ministry of Education, Culture, Sports, Science and Technology (MEXT), High Energy Accelerator Research Organization (KEK) and Japan Aerospace Exploration Agency (JAXA) in Japan, and the K.~A.~Wallenberg Foundation, the Swedish Research Council and the Swedish National Space Board in Sweden. 
 
Additional support for science analysis during the operations phase is gratefully acknowledged from the Istituto Nazionale di Astrofisica in Italy and the Centre National d'\'Etudes Spatiales in France.

GD and EG are members of CIC-CONICET (Argentina), LD is Fellow of CONICET (Argentina). They are supported through grants from CONICET and ANPCyT (Argentina). We acknowledge to Estela Reynoso and Anne Green who collaborated in the first stages of the HI data acquisition and preocessing.

%Our morphological analysis showed that the correlation between the GeV emission and the non-thermal X-ray signal is much better than with the thermal X-ray emission.

\bibliographystyle{apj}
\bibliography{rcw86_pass8}

\clearpage

\begin{table}[ht]
  \centering
  \caption{Values of the parameters for the different modelings of the broadband spectrum of RCW~86. \label{tab:models_parameters}}
  \begin{tabular}{|l|c|cc|cc|}
    \hline
    \hline
    \multirow{3}{*}{Parameters} & \multicolumn{3}{c}{Whole remnant} & \multicolumn{2}{|c|}{Regions}\\
     & One-zone & \multicolumn{2}{c|}{Two-zone} & \multirow{2}{*}{NE} & \multirow{2}{*}{SW}\rule[-3pt]{0pt}{12pt}\\
     & & Radio & X-ray & & \\
    \hline
    Density (cm$^{-3}$)                   & 0.1              & 1.0    & 1.0              & 1.0                    & 1.0\rule[-3pt]{0pt}{12pt}\\
    B ($\mu$G)                            & $10.2 \pm 0.5$   & $24$   & $10.5 \pm 0.7$   & $11.6\pm 0.7$          & $16.8 \pm 2.1$\\
    $\Gamma_{\rm e,p}$                    & $2.37 \pm 0.03$  & $2.2$  & $2.21 \pm 0.1$   & 2.21 & 2.21\\
    $E_{\rm max}$ (TeV)                   & $75 \pm 5$       & 2      & $67 \pm 4$       & $40 \pm 5$             & $61 \pm 5$\\
    $\eta_{\rm e}$ (\% of $E_{\rm SN}$)   & $3.84 \pm 0.5$   & 0.03   & $0.37 \pm 0.02$  & $0.14 \pm 0.04$        & $0.16 \pm 0.04$\\
    $\eta_{\rm p}$ (\% of $E_{\rm SN}$)   & 2                & -      & 2                & 0.5                    & 0.5\\
    K$_{\rm ep}$ ($\times$ 10$^{-2}$)     & $11.1 \pm 1.5$   & -      & $13.6 \pm 0.5$   & $5.2 \pm 1.5$          & $5.9 \pm 1.5$\\
    \hline
  \end{tabular}
\end{table}

\end{document}